# Unifying monitoring and modelling of water concentration levels in surface waters

*Peter B Sorensen*[*1], *Anders Nielsen*[2], *Peter E Holm*[5], *Poul L Bjerg*[4], *Denitza Voutchkova*[3], *Lærke Thorling*[3], *Dorte Rasmussen*[6], *Hans Estrup*[1], *Christian F Damgaard*[1]

1) Aarhus University, Institut for Ecoscience, Denmark

2) Danish Technical University, AQUA, Institute of Aquatic Resources, Denmark

3) Geological Survey of Denmark and Greenland, Department of Geochemistry, Denmark

4) Danish Technical University, Department of Environment and Resource Engineering, Denmark

5) University of Copenhagen, Department of Plant and Environmental Science, Denmark

6) Danish Hydraulic Institute (DHI), Denmark

[*]pbs@ecos.au.dk

## Abstract

Accurate prediction of expected concentrations is essential for effective catchment management, requiring both extensive monitoring and advanced modeling techniques. However, due to limitations in the equation solving capacity, the integration of monitoring and modeling has been suffering suboptimal statistical approaches. This limitation results in models that can only partially leverage monitoring data, thus being an obstacle for realistic uncertainty assessments by overlooking critical correlations between both measurements and model parameters. This study presents a novel solution that integrates catchment monitoring and a unified hieratical statistical catchment modeling that employs a log-normal distribution for residuals within a left-censored likelihood function to address measurements below detection limits. This enables the estimation of concentrations within sub-catchments in conjunction with a source/fate sub-catchment model and monitoring data. This approach is possible due to a model builder R package denoted RTMB. The proposed approach introduces a statistical paradigm based on a hierarchical structure, capable of accommodating heterogeneous sampling across various sampling locations and the authors suggest that this also will encourage further refinement of other existing modeling platforms within the scientific community to improve synergy with monitoring programs. The application of the method is demonstrated through an analysis of nickel concentrations in Danish surface waters.

# Introduction

The concentration of contaminants and nutrients in water plays a crucial role in determining the risk of adverse ecological effects, both within the catchment area and in the downstream marine environments receiving discharge from these areas. Accurate knowledge of expected concentrations is essential for effective catchment management practices, thereby needed for both monitoring and modeling efforts. Investments in the quantitative design of water quality monitoring programs, are therefore important to support investment strategy by introducing novel design methodologies that are linked to complex socio-economic activities, to establish a quantitative relationship between surface water measurements and emerging monitoring technologies (Jiang et al., 2020). The statistical approach employed in the present study can support such an investment strategy. It is not just a question of getting the maximal benefit of existing monitoring data by modelling, but also to be better of planning future monitoring activities by applying an optimal experimental design technique that integrates models and data collection (Li et al., 2023) and this also needs a combination of modeling and data collection using a fully integrated statistical model.

Catchment models are typically constructed based on downstream transport mass balance equations and have been developed during the last three decades, as e.g. the SWAT model (Arnold et al., 1998), GREAT-ER (Koormann et al., 2006), STREAM-EU (Lindim et al., 2015), ePie (Oldenkamp et al., 2018), and also more specific and thus complex models as for metal pollution transport (Sui et al., 2022) or more general multiple media models as GEM (Little et al., 2018). All these models are developed for management purposes and integrate process modeling including substance transfer between different medias and degradation, with a water flow model (encompassing both groundwater and surface water), using downstream transported mass balance equations to predict concentration levels. However, the linkage between catchment monitoring and modeling has historically been limited, often relying on what is denoted as invers modelling (Boxall et al., 2014) or residual based modelling (Guo et al., 2022), which basically is a model calibration applying calibration methods more complex than a simple classical regression but they are applying a fully integration of the statistics of the monitoring data. An evaluation tool for SWAT in form of R-SWAT is highly developed (Nguyen et al., 2022), but is still lacking the advances of more complete statistical methods. The lack of full synergy between catchment monitoring and modeling can be attributed to the mathematical challenges of solving likelihood function equations that account for the complexity of the multi-parameter space defined by catchment models. The challenge arise in the statistical model, where the likelihood function needs to be integrated over and interval of all realistic values in the a full parameter space having dimensions equal the number of parameters, and as a catchment model typically generates a parameter space in thousands of dimensions yielding a huge parameter space for integrating the likelihood function. This has been the obstacle for full application of statical models, but an equation solver has now been introduced within the R software framework, capable of handling the full statistical complexity inherent in catchment models, based on the TMB solver (Kristensen et al., 2016). As a result, it is now realistic to unify catchment monitoring and catchment modeling by a fully statical modelling approach.

It is fair to say that there still is a way to go beyond this paper for finishing a full synergy between a catchment modelling and a complete statistical modelling, but this paper is opening the game for the next generation of such models by solving fundamental obstacle for the synergy by introducing new strong equation solver in R to catchment modelers, based on the equation solver TMB (K. Kristensen et al., 2016). Statistical modelling of stream networks has been applied by for experimental planning of monitoring activities, using a so-called tail up and tail down approach for estimating the spatial correlation of measurements due to substance mass transport caried by the water flow direction within the stream network (Som et al., 2014). In this paper, the approach of up and down tailing, is not needed for the model as the equation solver is strong enough to solve the system of mass balance equations directly, so the data set used for testing in this paper is a full national wide data set from the monitoring program of Denmark for nickel.

The integrated modelling concept combines two components: (1) a model that incorporates measurements to estimate expected concentrations within each sub-catchment; and (2) a source/fate model that utilizes catchment information to predict expected mass input and removal for each sub-catchment. A hydraulic model serves as input for both the measurement and source/fate models, while additional relevant catchment data is exclusively utilized by the source/fate model. The source/fate model can take many forms and thus be based on selected existing approaches already applied in catchment modeling, and the purpose of this paper is to propose a novel statistical paradigm that can inspire and guide the modification of other existing or future catchment modeling tools, with the aim of enhancing the integration and application of monitoring data.

## Principle

The principle is based on the division of catchment areas into sub-catchments defined as a region where all water discharge ultimately converges downstream to a single outlet that empties into the marine environment. A sub-catchment refers to a specific sub area within the catchment, where all discharge is assumed drained to rivers/streams that flow within the sub-catchment's boundaries. It thus represents a specific part of the catchment where water drains into a smaller stream or tributary before joining the main water body.

The model concept integrates four models:

- *Catchment model*, where the model area is divided into catchments and sub-catchments using the topographic catchment and the water flow between sub-catchments is assumed known from hydraulic measurements and/or models.
- *Measurement model*, where the measurements are assumed to be representative of a sub-catchment outlet following a log-normal distribution.
- *Source model*, where a mass source (emission) model is parameterized and used for predictions
- *Removal model*, where mass removal model (retention and/or degradation) is parameterized and used for predictions

The source and removal model are hypothetical because the validity of these models is tested using the statistical model. Such testing may lead to replacement or improvement of the source/removal models. The conceptual framework is visually represented in Figure 1, showing a single catchment divided into sub-catchments.

The input data needed for the source/removal models are gathered for each sub-catchment. The source/removal models can either be simple or complex and based on existing model principles used in catchment modeling, by relating properties in the sub-catchment to respectively the source and removal of substance. The input from monitoring data is the measured concentration for some sub-catchment outlets, as illustrated in Figure 1 as five red marked sub-catchment outlets. Thus, only a fraction of the sub-catchments will be sampled at the outlet. The modeling principle is based on a mass balance approach for each sub-catchment, where the mass of a given substance is delivered into the sub-catchment from upstream sub-catchments. To this mass is added the sources within the sub-catchment as described by the source model and subtracted the removed mass as described by the removal model. The mass balance of each sub-catchment is fitted to the measured concentration using the likelihood function in the statistical model. The output is predicted outlet concentration of each sub-catchment and value estimates for all parameters used in the source/ removal models. All estimates are distribution functions, thus having an individual standard deviation.

The main output contains:

- Source/removal model parameter estimation. The source/removal models will always include parameters that needs calibration. These parameters are estimated using the statistical model as distributions that quantify the uncertainty of the estimates:
    - Source model parameters. This includes all the needed parameters to feed the source model. It could e.g. be an emission factor to multiply to agricultural areas for getting mass flow to surface waters and that could be adjusted by a macropore index for the soil in the catchment. In this paper for the two-source version, the source model will be a groundwater inflow model.
    - Removal model parameters. This will typically include parameters such as the half lifetime of a substance or a retention factor. In this paper the removal model is a first order retention model for both the one source and two source version.
- Unexplained variation by the Source/removal model estimated by the statistical model. This estimation divides the unexplained variation into:
    - Replicant variation, which is the variation between single measurements for the same sub-catchment outlet. This type of variation can be considered as random noise and thus by definition a variation that are out of reach of the source/removal models to be explained.
    - Unexplained variation between sub-catchments. In the ideal case, the source/ removal models will describe all the variation between sub-catchments, but there will always be variation between sub-catchments that are left back as unexplained by the source/removal models.
    - Unexplained variation between years. This type of variation arises from different conditions between years for the whole modelling area. It could be due to a change in regulations or difference in climate. However, it could also be associated with a heterogeny strategy of the monitoring design of sample collection and in this case the estimation of variation between years is useful for statistical elimination of artifacts arising from monitoring design that may bias the parameter estimation.

- Predicted outlet concentration. The source/removal models integrated in the statistical model can predict the outlet concentration for all sub-catchments. The prediction will be concentration level distributions having an expected value and a standard error.

In the visual description in Figure 1, samples for concentration measurements are collected at five specific catchment outlets (red circles), with all other outlets remaining unsampled (Green circles). The statistical model can predict expected concentrations both for sampled and unsampled outlets, but the uncertainty will typically depend on whatever the outlet is sampled or not. The Unsampled outlets are connected to the sampled outlet by the mass balance between sub-catchments, where the mass transport at a sampled outlet needs to be reflected in connected unsampled outlets. Thus, the measured concentration values are used to predict expected concentrations at all sub-catchment outlets included in the same catchment, also for the unsampled outlets.

The mass balance for each sub-catchment also includes models that predict source of mass into surface waters, as well as mass removal due to degradation and/or sedimentation. The novel aspect of this study is the full integration of the statistical rules and mass balance models for a larger system of catchments and sub-catchments. This approach also involves estimating all model parameters and their associated standard deviations, providing a comprehensive framework for mass transfer and concentration estimation within the entire water system. The mathematical details are described below under Method.

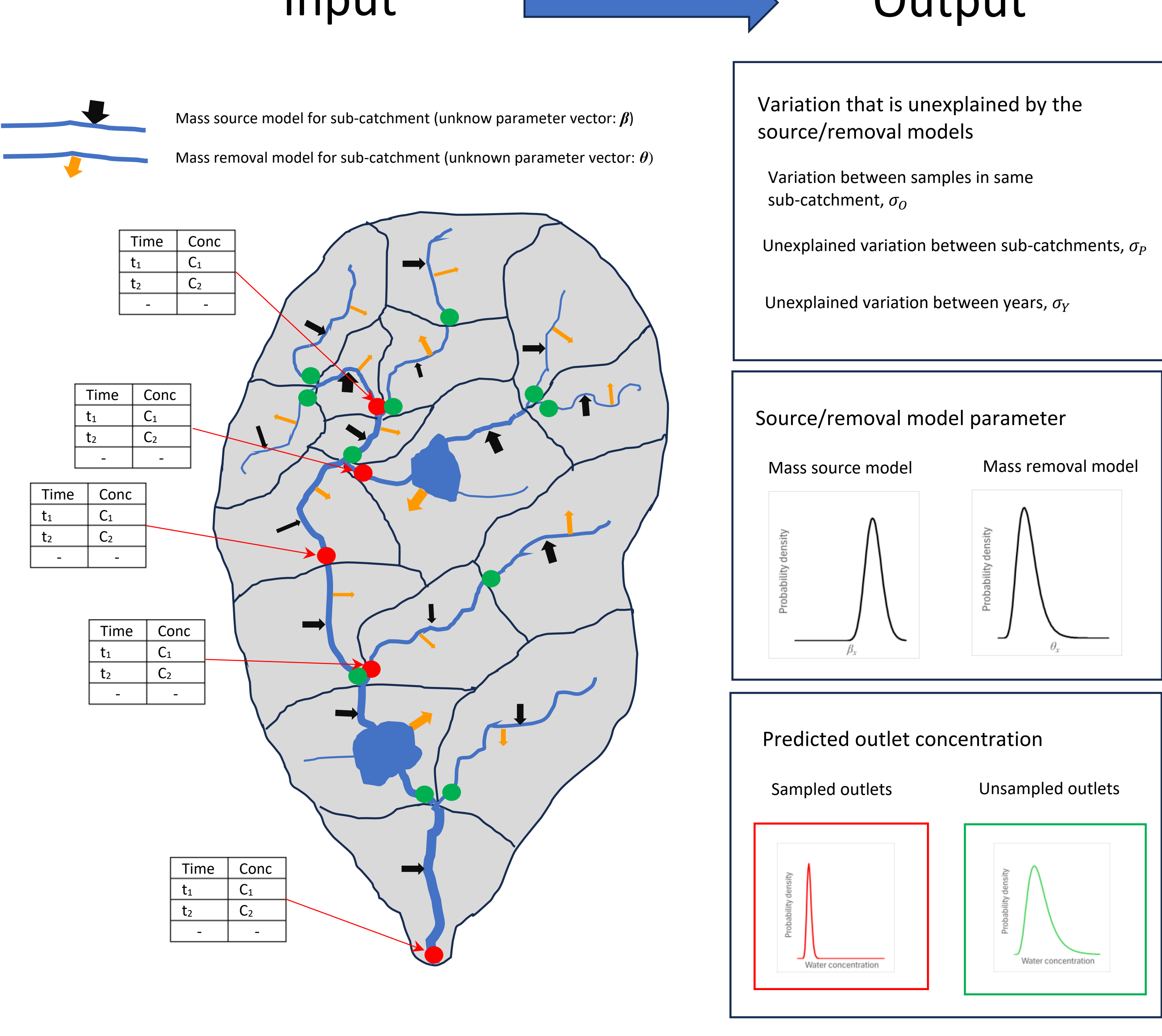

Figure 1. Visual description og the modeling principle where a river catchment is divided into sub-catchments, each characterized by an expected concentration at the outlet. A subset of these outlets has been sampled for water concentration measurements, shown as red circles, while green circles represent unsampled sub-catchment outlets. The catchment encompasses both rivers/streams and lakes, with black arrows indicating source of mass into surface waters within the sub-catchment, and orange arrows representing corresponding mass removal due to degradation and/or sedimentation processes. The thickness of the arrows corresponds to varying rates of mass transfer.

# Method

## Catchment model

A series of non-overlapping sub-catchment areas are defined as SubCat1, SubCat2, ..., in such a way that water can flow from one sub-catchment to the next through a single outlet point. The sub-catchment hierarchy is defined as illustrated in Figure 2, where sub-catchments with no upstream sub-catchments are assigned a level of 0. Sub-catchments that receive input solely from level 0 sub-catchments are assigned to level 1, and this hierarchy continues accordingly for higher levels.

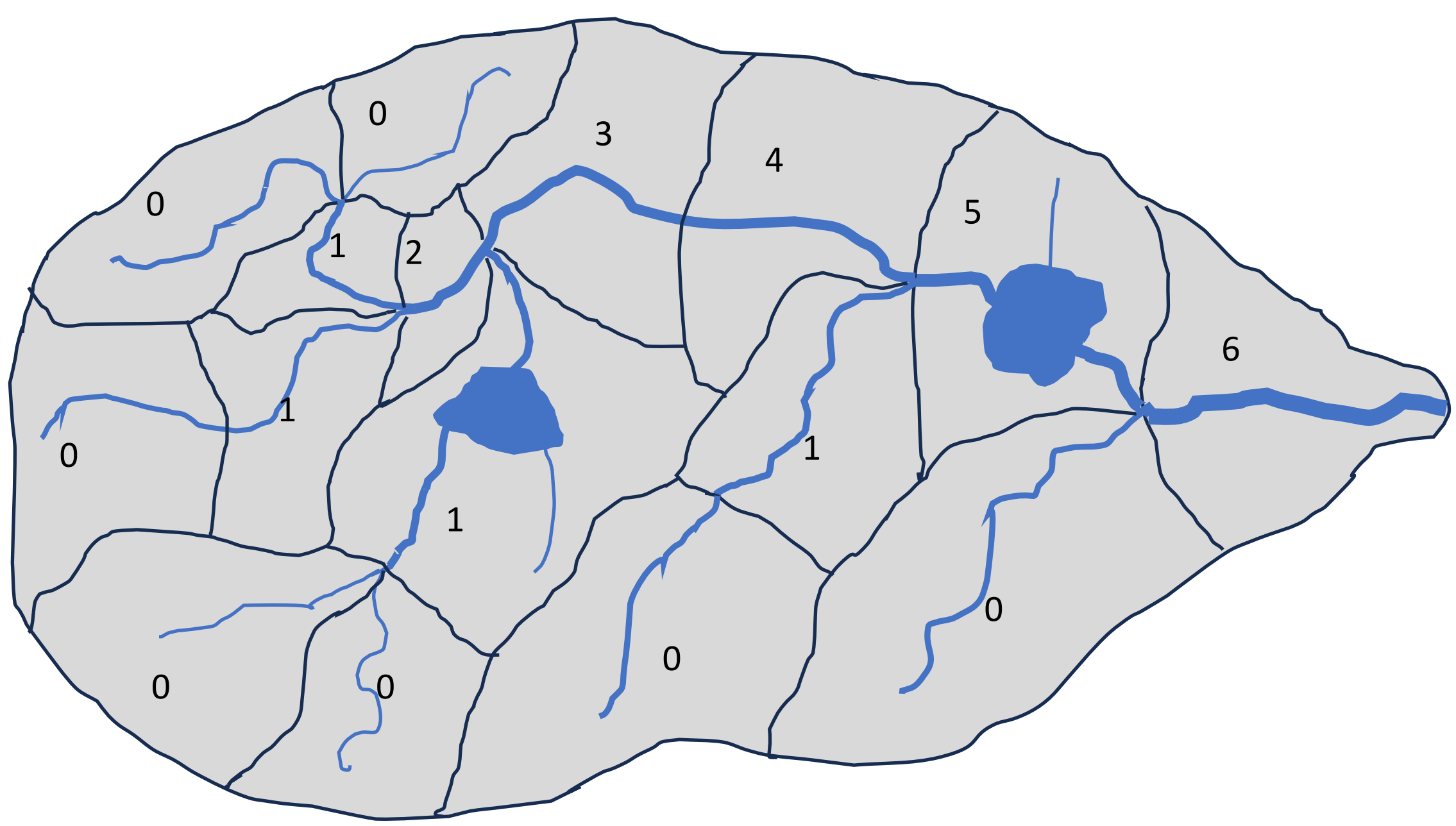

Figure 2. Definition of catchment levels based on the longest upstream chain, where level 0 corresponds to catchments that have no upstream catchments.

The catchment levels are utilized to perform a stratified calculation, where the expected concentration is first calculated for all level 0 catchments, followed by the calculation for all level 1 catchments, and so on, until the expected concentration is determined for all sub-catchments. This procedure ensures that the mass balance for each substance is always defined for a sub-catchment when the expected concentration is computed first.

As depicted in Figure 2, each sub-catchment has only one outlet, but it may receive water from none or multiple upstream catchments. Each sub-catchment discharges water either to the marine environment or to a single downstream sub-catchment. This information is mathematically organized by this definition:

$$\Psi_{i,j} \equiv \begin{cases} 1\text{, if } SubCat_i \xrightarrow{discharges} SubCat_j \\ 0\text{, otherwise} \end{cases} \qquad (1).$$

Where $\mathbf{\Psi}$ is the sparse connectivity matrix. Thus, if sub-catchment $i$ is discharging into a marine area then $\sum_{j=1}^{N} \Psi_{i,j} = 0$, where $N$ is the number of sub-catchments.

The accumulation of mass is assumed negligible in the mass balance for each sub-catchment (steady state) yielding the following mass balance of the substance:

$$M_{out} = M_{in} - M_{ret} - M_{deg} \quad (2),$$

The following hydrologic parameters are assumed known, having fixed value for each sub-catchment:

- The total water discharge from the sub-catchment ($Q_i$)
- The amount of groundwater entering surface waters in the sub-catchment ($Qg_i$)
- The amount of shallow water including surface runoff and tile drain flow entering surface waters in the sub-catchment ($Qs_i$)
- The total surface water area within the sub-catchmen ($A_i$)

The water flow components are calculated using the DK model (Liu et al., 2024) and the surface water area is calculated based on a GIS analysis using  (Levin, 2017). The mass balance of water is assumed to be steady as:

$$Q_i = \sum_{j=1}^{N} Q_j \cdot \Psi_{i,j} + Qg_i + Qs_i \quad (3)$$

Thus, the governing hydraulic assumptions is that there is no accumulation of water in the catchment, no evaporation/precipitation, and the flow is assumed constant.

## Mass balance for a sub-catchment

### Non steady state equation

Each sub-catchments are assumed to be completely mixed, yielding the mass balance for a surface water transported substance:

$$\frac{d(C_i V_i)}{dt} = \sum_{j=1}^{N} C_j Q_j \Psi_{i,j} + f(\boldsymbol{\beta}, \boldsymbol{X_i}) - Q_i C_i - g(\boldsymbol{\theta}, \boldsymbol{R}_i, C_i) \quad (4),$$

where $C_i$ is the expected concentration in sub-catchment $i$, $V_i$ is the water volume in the sub-catchment $i$, $Q_j$ and $Q_i$ are the flow of water out of respectively sub-catchment $j$ and $i$. The function $f(\boldsymbol{\beta}, \boldsymbol{X_i})$ is the source model for sub-catchment $i$, where $\boldsymbol{\beta}$ is a parameter vector that will be estimated by the model and $\boldsymbol{X_i}$ is the corresponding vector of independent variables for sub-catchment $i$, that are used as input to the source model. The function $g(\boldsymbol{\theta}, \boldsymbol{R}_i, C_i)$ is the removal model, in where $\boldsymbol{\theta}$ is the parameter vector that will be estimated by the model and $\boldsymbol{R}_i$ is the vector of independent variables for sub-catchment $i$ that are used as input to the removal model. This paper applies the following version of the Eq. 4:

$$\frac{d(C_i V_i)}{dt} = \sum_{j=1}^{N} C_j Q_j \Psi_{i,j} + \boldsymbol{\beta} \boldsymbol{X_i} - Q_i C_i - \theta A_i C_i \quad (5),$$

Where the source model is the product between the parameter vector $\boldsymbol{\beta}$ and the variable vector $\boldsymbol{X_i}$ for the catchment $i$, $\theta$ is the retention coefficient (single element vector) and $A_i$ is the surface water area in sub-catchment $i$. The removal model in Eq. 4 is "$\theta A_i C_i$", where the retention is assumed to be first order in relation to surface water area and concentration level.

The argument for this relation is that the relation is typically adsorption and degradation on contact surface of macrophytes or sediments, or a sedimentation out of water column that will not be complete before the water element enters the outlet. Nevertheless, other source and removal models could have been chosen as well, and it would be possible to apply the statistical model to select the best source/removal model out of several alternatives.

### Steady state equation

The mass balance model applied in the paper is a simple steady state model assuming no accumulation of substance in the sub-catchment. This means that the contamination history for the substance is assumed long enough for stabilizing toward constant concentration levels in the sub-catchment and the Eq.5 is simplified as

$$\frac{d(C_i V)}{dt} \equiv 0 \rightarrow C_i = \frac{\sum_{j=1}^{N} C_j \cdot Q_j \cdot \Psi_{i,j} + \boldsymbol{\beta X_i}}{Q_i + A_i \cdot \theta} \qquad (6)$$

Every source is described in the catchment by an *X* value, where one source could be the groundwater contribution to the total water discharge, another one could be wastewater from treatment plant facilities in the catchment or agricultural area that potentially could contribute with mass from agricultural activity, thus forming a vector $\boldsymbol{X}$ for each sub-catchment. All those factors are potential contributions, with unknown real impact on the mass balance for the catchment, so the $\boldsymbol{X}$ vector is weighted with a $\boldsymbol{\beta}$ vector having a coefficient for each source which assumed similar in value for all catchments. The value of $\boldsymbol{\beta}$ is estimated including uncertainty estimates. So, the mass transport to surface waters in sub-catchment *i* is:

$$M_i = \boldsymbol{\beta X}_i \,, \;\; \beta_i \geq 0 \qquad (7),$$

where all elements in vector $\boldsymbol{\beta}$ are positive and the values will be estimated by the model. As a log-normal model is used for estimation, the intercept and effects of e.g. years in the statistical model will interfere with the estimation of the $\boldsymbol{\beta}$ vector. To solve this challenge, the first element of $\boldsymbol{X}$ ($X_{i,0}$) is defined to be the factor of reference as

$$log(\boldsymbol{\beta X_i}) = log(\beta_0) + log\left(X_{i,0} + \sum_{k=1}^{P} \beta'_k X_{i,k}\right), where \;\; \beta_k = \frac{\beta'_k}{\beta_0} \;, \; for\; k = 1,2..,M \qquad (8),$$

where *M* is the number of factors in the source model. Selecting the reference factor and thereby $\beta_0$ is a matter of definition, however, it must be a factor that actually will contribute to the source to some extent. The selection of the reference is illustrated and discussed below in the case study of nickel in Danish surface waters.

Combining Eq. 6 with the non-log of Eq. 8 yields:

$$C_i = \frac{\beta_0 \left( \frac{\sum_{j=1}^{N} C_j \cdot Q_j \cdot \Psi_{i,j}}{\beta_0} + X_{i,0} + \sum_{k=1}^{P} \beta'_k X_{i,k} \right)}{Q_i (1 + \theta \cdot \frac{A_i}{Q_i})} \qquad (9)$$

The log form of this equation is combined with the measurement model below.

### Measurement model

The Eq. 9 is applied for calculating the log-concentration as:

$$lC_i = \log\left(\frac{\sum_{j=1}^{N} C_j \cdot Q_j \cdot \Psi_{i,j}}{\beta_0} + X_{i,0} + \sum_{k=1}^{K} \beta'_k \, X_{i,k}\right) - \log(Q_i) + \log(\beta_0) - \log\left(1 + \theta \, \frac{A_i}{Q_i}\right) + \varepsilon_i \qquad (10)$$

Where $lC_i = log(C_i)$, and $\varepsilon_i$ is the model residual assumed distributed as $\varepsilon_i \sim N(0, \sigma_P^2)$ and where $\sigma_P^2$ is the variance of residuals, which can be seen as the variance between sub-catchments of the expected concentration that have not been explained by the source/removal modelling. It is important to make clear that the residual $\varepsilon_i$ is not the residual between single measurement of samples, but instead the residual between expected concentration and modeled concentration by the source/removal models.

The measured concentrations in the surface waters are left-censored at a known detection limit for each sample ($D_{i,y,j}$) and assumed to follow a truncated log-normal distribution, so in the case a measurement is below the detection limit, it is modelled by the cumulative log-normal distribution. The detection limit is allowed to vary between samples. Furthermore, the concentrations are assumed to vary from year-to-year at the national level. The unknown but true mean values of the log-transformed concentrations at sub-catchment *i* are in this way modelled statistically using the measured concentrations at the sampling sites are indexed as $z_{i,y,j}$, where *y* is the sampling year, and *j* are replicates from the same area and year. Consequently, the likelihood function of the measurement equation is

$$\mathrm{L}(z_{i,y,j}) = \begin{cases} \phi(\log(z_{i,y,j}), lC_i + \delta_y, \sigma_O^2), & z_{i,y,j} \geq D_{i,y,j} \\ \Phi(\log(D_{i,y,j}), lC_i + \delta_y, \sigma_O^2), & z_{i,y,j} < D_{i,y,j} \end{cases} \qquad (11),$$

where $\phi(\log(z_{i,y,j}), lC_i + \delta_y, \sigma_O^2)$ and $\Phi(\log(D_{i,y,j}), lC_i + \delta_y, \sigma_O^2)$ are density and cumulative density respectively for the normal distribution with mean $lC_i + \delta_y$ and variance $\sigma_O^2$ evaluated in respectively $\log(z_{i,y,j})$ for $\phi$ and $\log(D_{i,y,j})$ for $\Phi$ and $\delta_y$ are random effects of year with density function $\phi(0, \sigma_Y^2)$. The variance $\sigma_O^2$ is the variance between repeated measurements (replicants) sampled at the same sub-catchment, so $\sigma_0$ is the standard deviation of the replicants.

It is a major benefit of a full statistical modelling that we have established the two variance components $\sigma_O^2$ and $\sigma_P^2$ as they can evaluate the modelling of the entire system. The variance $\sigma_O^2$ quantifies the variation that cannot be explained by "any model" for the chosen structure of sub-catchments, as this is random noise between single measurements for the same sub-catchment. Only, structural change of the catchment model setup as a more fine-grained sub-catchment division and/or better inclusion of time effects could limit $\sigma_O^2$. So, the variance $\sigma_P^2$ quantifies unexplained variation between sub-catchments and thus variation that potentially can be explained by refinement of the source/removal models to better separate the different conditions of the sub-catchments.

## Estimation

The predicted concentration (Eq. 10) depends on the predicted concentration of the upstream catchments that discharge to the sub-catchment. To handle this dependency, the Eq. 10 is initially calculated for all level 0 sub-catchment where $\sum_{j=1}^{N} C_j \cdot Q_j \cdot \Psi_{i,j} = 0$ and then only for level one catchment, followed by all level two catchments, etc.

The statistical model (Eqs. 10 and 11) are fitted to measured concentration data using the R software package RTMB integrating the R environment with TMB modelling framework (Kasper Kristensen et al., 2016), in which automatic differentiation and the Laplace approximation enables very fast estimation of the parameters and latent sub-catchment concentrations in the model.

## Results

Application of two simple versions of the model are discussed using data from the nation-wide environmental monitoring data for nickel in Denmark (Miljøstyrelsen, 2017). The country is divided into 3350 sub-catchments and 532 (16 %) of these have sampling points close the sub-catchment outlet and is thus sampled leaving 84 % as unsampled. There are totally 6018 single samples (filtered samples) that are taken nationwide during the period from 2011 to 2023, and out of these 55 are measured under the detection limit. The surface water area ($A_i$) is estimated by matching the shape files of the sub-catchments with data from BaseMap02 (Gregor L., 2017).

For illustration of the measurement model, then first application is a simple single source analysis, where the only source of nickel is assumed to be a constant nationwide concentration level in the inflow of water to the surface water. The single source version seems to be very restricted and thus limited model, and of cause it is, but, there will come very useful results showing the modelling potential from this analysis.

The single source analysis will be followed by an extension to a two-source version by splitting the source up into two components as groundwater born sources and other sources. This will split the source of nickel into two main categories as a groundwater born source primary from natural mineral origin on one side and a more surface-near source on the other side.

### Single source version

As a first step the only source of mass is simply the nationwide distributed concentration of nickel in generated water discharged to the sub-catchments ($Qg_i + Qs_i$). So, this is a version all discharge to the ID15 catchments is assumed to have the same nickel concentration and $X_{i,0}$ will then equals the inflowing amount of water to each catchment:

$$X_{i,0} = Qg_i + Qs_i \tag{12}$$

and the predicted concentration equation becomes:

$$lC_i = \log\left(\frac{\sum_{j=1}^{N} C_j \cdot Q_j \cdot \Psi_{i,j}}{\beta_0} + X_{i,0}\right) - \log(Q_i) + \log(\beta_0) - \log\left(1 + \theta\ \frac{A_i}{Q_i}\right) + \varepsilon_i \tag{13}$$

The intercept ($\beta_0$) is thus the estimated nationwide mean nickel concentration of the generated discharge to the sub-catchments ($Qg_i + Qs_i$).

The general parameter estimation is seen in Table 1, in columns for Singl-source. The standard deviation between sub-catchments is 0.563 indicating that there is a considerable variation between sub-catchments due to different conditions. The replication standard deviation of 0.440 is a little lower, but still considerable, while the standard deviation between years is limited having the value of 0.121. The retention factor is estimated to be 35.6 and the importance for the overall mass balance will be discussed later as part of a country scale assessment. The national wide mean concentration of the inflowing water is 1.45 µg/l. All estimates have assigned a standard error estimate (SE) and the T factor is defined as the ratio between Estimate and SE, indicating the strength of the statement that the estimate is significantly different from zero.

| | Single-source (Eq. 13) | | | Two-source (Eq. 15) | | |
|---|---|---|---|---|---|---|
| Parameter | Estimate | SE | T | Estimate | SE | T |
| $\sigma_0$ | 0.440 | 0.004 | 109.8 | 0.440 | 0.004 | 109.8 |
| $\sigma_P$ | 0.563 | 0.019 | 5.1 | 0.513 | 0.017 | 5.1 |
| $\sigma_Y$ | 0.121 | 0.024 | 30.1 | 0.124 | 0.024 | 29.4 |
| $\theta$ | 35.6 | 9.3 | 3.8 | 25.28 | 6.533 | 3.9 |
| $\beta_0$ [µg/l] | 1.45 | 0.078 | 18.6 | 1.75 | 0.121 | 14.4 |
| $\beta_1'$ [l/µg] | - | - | - | 0.529 | 0.088 | 6.0 |
| $\beta_1 = \beta_0\beta_1'$ | - | - | - | 0.93 | - | - |

Table 1. The general output for nickel in surface waters at national level in Denmark and from respectively the single and two source version of modelling. $\sigma_0$ is replicant standard deviation, $\sigma_P$ is standard deviation between sub-catchments, $\sigma_Y$ is standard deviation between years, $\theta$ is retention factor, $\beta_0$ is reference (intercept), $\beta_1'$ is parameter for ground water source, $\beta_1$ is scaling factor for ground water source model to be correct.

The general model fit for the single-source model is displayed in Figure 3 as a scatter plot having the predicted log-concentration on the x-axis and the corresponding measured log-concentration on the y axis. The *y=x* relation is displayed as the red line and separation of measurements along this line is a graphical display of the difference between the sub-catchments as described by $\sigma_P$, while the random variation along the line illustrates the standard deviation of the replicates ($\sigma_0$). If the *y=x* line does not follow the center of the point cloud from the smallest to largest predictions, then the relation $\varepsilon_i \sim N(0, \sigma_P^2)$ is invalid for all sub-catchments, however, the line *y=x* seems visually close to the center in general for the hole interval. Larger positive data point deviation from the *y=x* line, especially for higher measured concentration levels are important to consult manually. In this case there are single measures that deviate from the expected values (*y=x* line), which could be related to an important event such as extreme weather, a station that is not representative for the outlet of the sub-catchment or simply an error in the monitoring database.

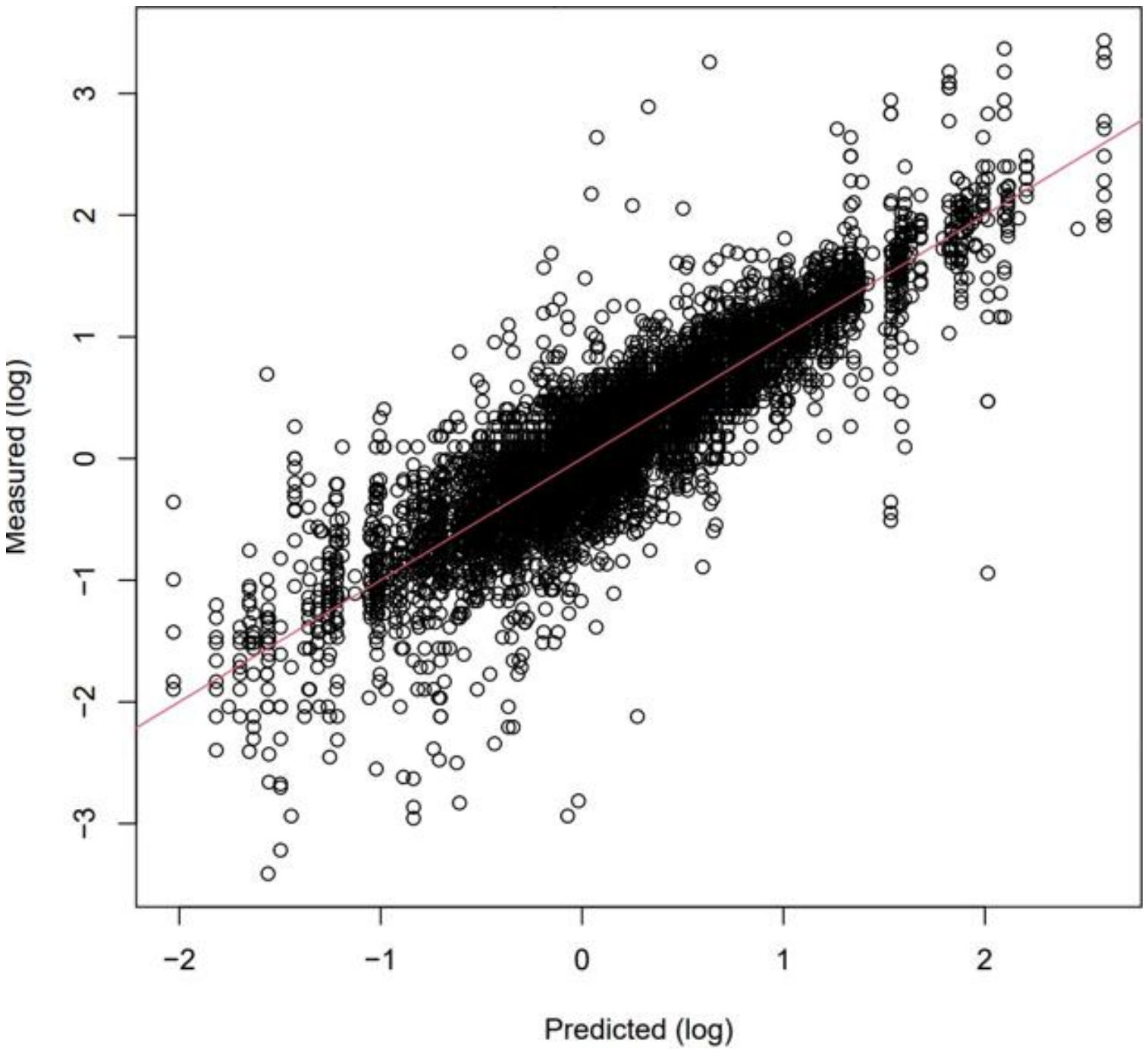

Figure 3. Scatter plot showing the national wide nickel concentration in Danish surface waters. The *x*-axis is the single source version of modelling and the y-axis displays measured concentration. The reed line is displaying the Measured=Predicted relationship. The model results are displayed on a map of all catchments on Figure 5, where the color represents values for the expected concentration, along with a graph showing the nationwide relative differences between years ($\delta_y$), with the 95% confidence interval indicated as $\pm 1.96 \cdot \sigma_Y$. Effects between years are estimated between monitoring stations, even if each station has not been measured every year. A slight tendency for a decline in nickel concentrations over the years is observed, with a zigzag pattern between 2020, 2021, 2022, and 2023. Such fluctuations could have a genuine explanation, such as differences between wet and dry years. Therefore, temporal trends should be considered over a longer period. However, if there is a general difference in the criteria for how stations are selected from one year to the next, the model will interpret this as an effect among years. Temporal effects are not only included in the model to investigate time trends, but also to eliminate potential differences between years, both artifacts and real ones, that could introduce bias into the model's other estimates.

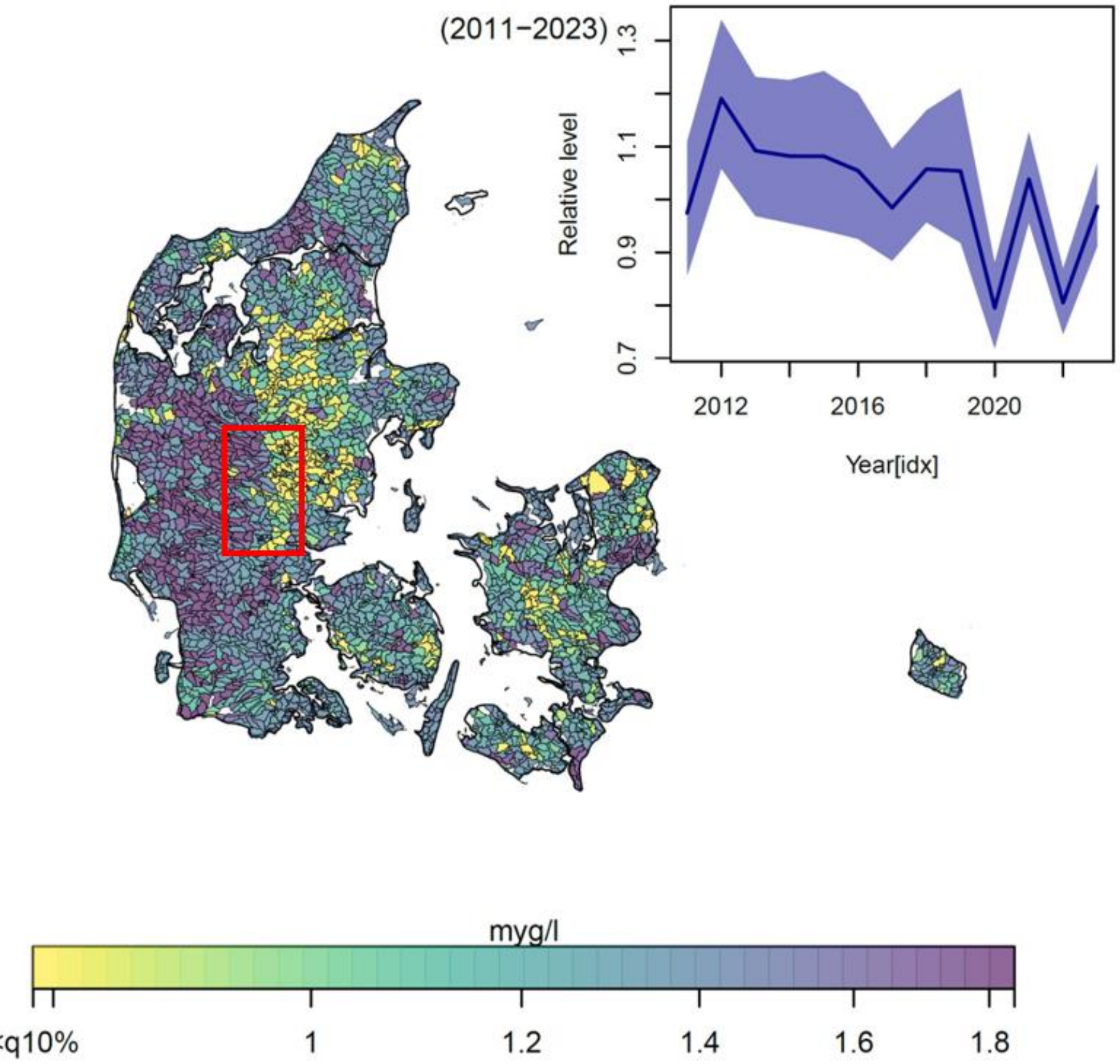

Figure 5. Predicted concentration (Eq.13) of all sub-catchments for nickel by the single source version of StatSurf and effect of year for each year reported in the data set and as relative difference between years. The reed frame shows the zoomed area for Figure 6.

It is evident that Figure 5 displays significant geographic differences in the expected concentrations, particularly along a south north edge in the western part of the country around the ice glacial margin from the last Ice Age. Such a clear geographic edge of differences is amplified as the model incorporates water exchange into the estimation of expected concentrations. This is illustrated in Figure 6, showing the boundary between two separated catchments, where the sampling stations are shown as circles, with the mean value of the measured concentrations at each station determining the color of the circle, according to the same color code applied to the sub-catchments. Additionally, river courses are overlaid with a line thickness corresponding to the hierarchy of each river section, as measured by the Strahler hierarchy (Horton, 1945), which distinguishes river sections hierarchically based on the greatest number of branches in the river network leading to the catchment boundary of the main catchment. Most monitoring stations are located near the outlet of a sub-catchment, and there is a clear tendency for the color of a circle to match the color of the expected concentration of the catchment where the station is located. This alignment is not surprising, as the model uses data both to determine the mean values in the circles and to estimate the expected concentrations in the sub-catchments. However, what is seen in Figure 6 is that also sub-catchments having no measurements are influenced by the measured catchment due to the water flow, and this effect generates a general division of sub-catchment areas of respectively high and low concentration.

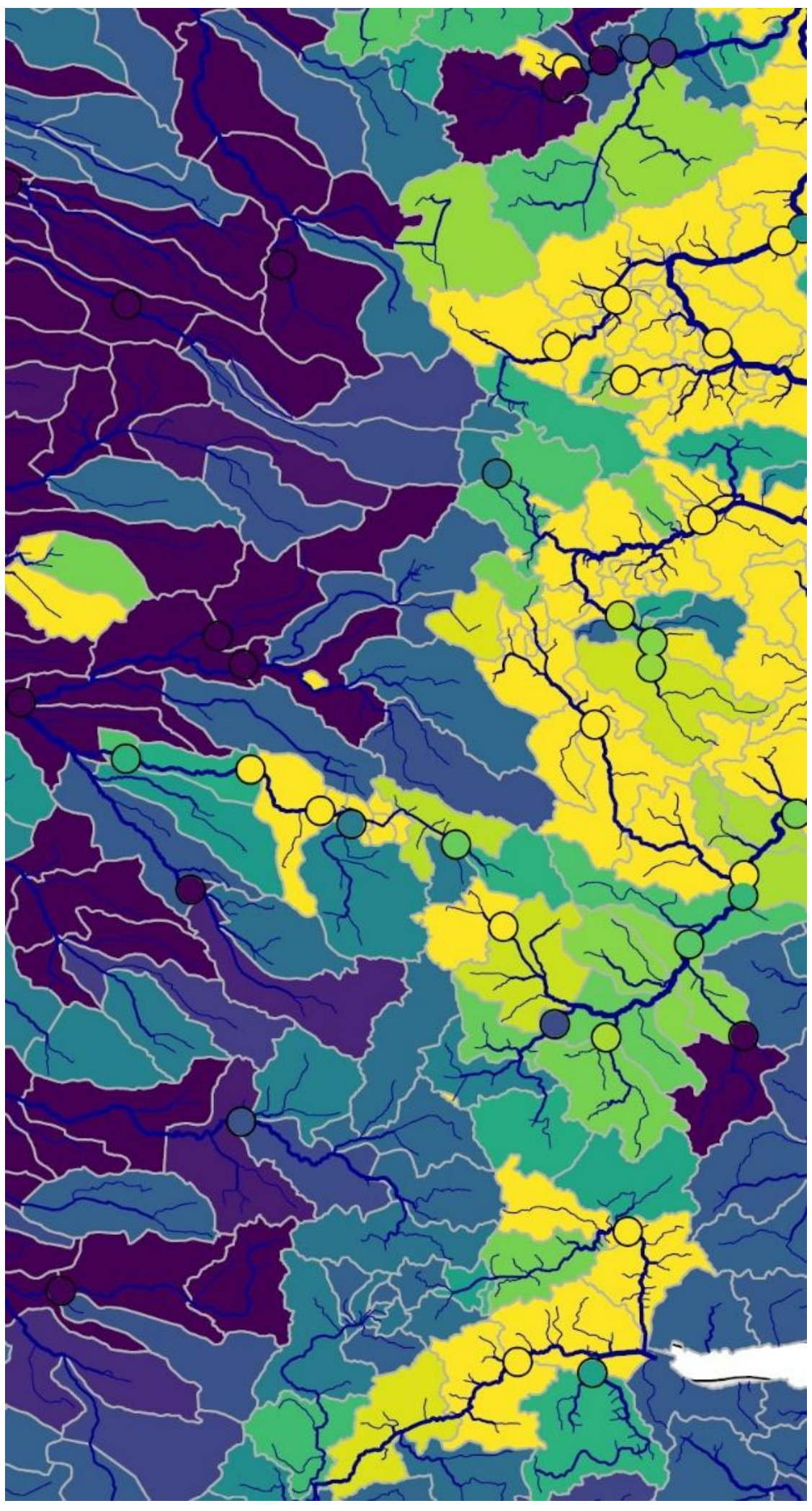

Figure 6. Showing the red frame in Figure 5 (Vest: Skjern Å and east: Guden Å), where the sampling stations are shown as circles, with the mean value of the measured concentrations at each station determining the color of the circle, according to the same color code applied to the catchments.

The clear division observed in Figure 6 between high and low concentration sub-catchments is further illustrated in Figure 7, for a series of connected sub-catchments (labeled: A, B, C, D, E, F). Catchment A is the downstream sub-catchment that has a sampling point located downstream at the outlet of catchment A, where high nickel concentrations are measured, corresponding to the dark colored circle, and consequently, sub-catchment A is assigned to a high predicted concentration. Much of the water flowing out of catchment A originates from sub-catchment B, therefore, the expected concentration from the un-measured catchment B is also predicted high, as it is essentially the same water being measured in sub-catchment A and, and the same holds also true for sub-catchment C. However, sub-catchment E is located far from sub-catchment A, so the water volume from this sub-catchment is modest compared to the larger outflow from sub-catchment A. This means that the expected concentration for sub-catchment E is facing more flexibility to be either low or high under the condition of high concentration in sub-catchment A. Consequently, the model will predict a concentration for sub-catchment E that is closer to what can be expected at the national level, without taking the measurements at sub-catchment A into account, and this tendency is further emphasized by the decreasing water volume up to sub-catchment F.

The outlet of sub-catchment D is a smaller stream, so also in this case there will be more flexibility in relation to the measurements in sub-catchment A. The single-source version model is conservative so predictions that deviates from the expected national values are only made when measurements necessitate this, so distant sub-catchments from sampling points will tend to be predicted as country means.

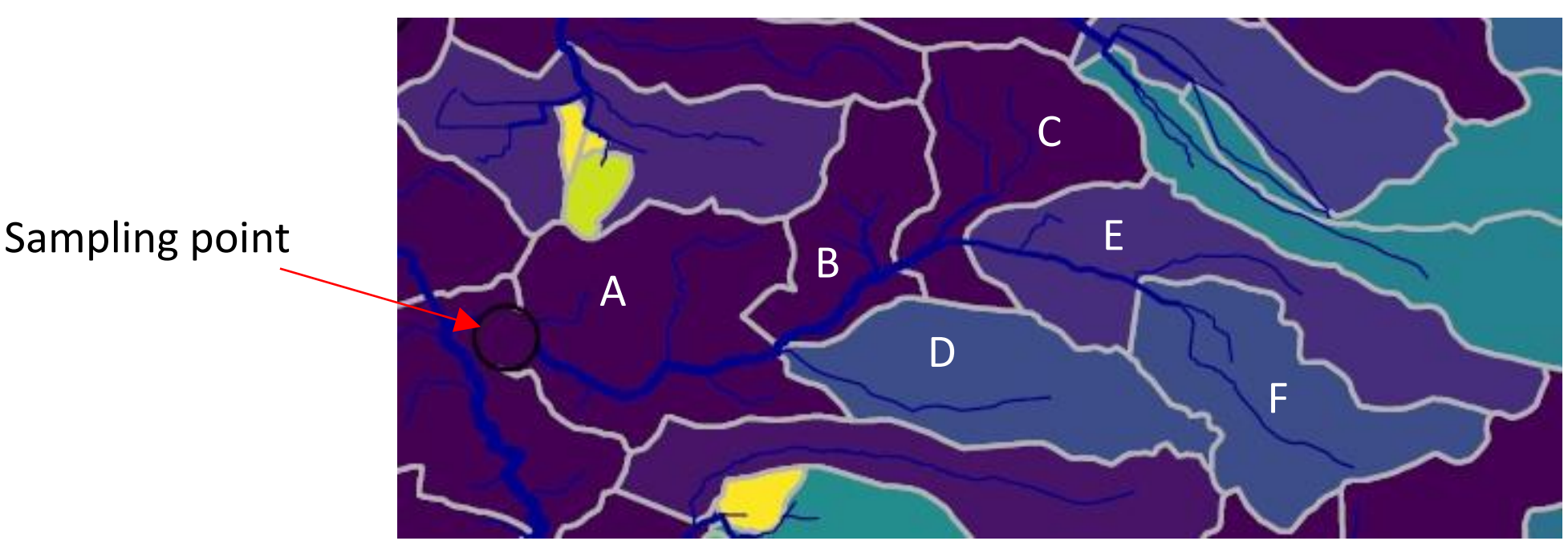

Figure 7. A series of connected sub-catchments (labeled: A, B, C, D, E, F). Catchment A is the downstream sub-catchment that has a sampling point located downstream at the outlet of catchment A. The color schemes are like Figure 5.

The above analysis focuses on the expected concentration in each catchment, which is one of the main outcomes of the model. However, the model can also estimate other useful variables. For example, the fitted model can be used to estimate the mass of nickel by multiplying the water volume flowing from the sub-catchments with the predicted concentrations, as shown in Figure 8. The figure clearly illustrates that the transported mass of nickel is dominated by the large rivers of Western Jutland, as these areas exhibit the highest concentrations, while the rivers also have high discharge. It is possible to calculate the total mass of nickel that is washed out to marine areas daily, amounting to 371 kg/day, and the total removal of nickel from surface waters by retention on the way to the marine environment can be estimated as 16 kg/day. The total mass entering the surface water is thus the sum of these two components equal 387 kg/day.

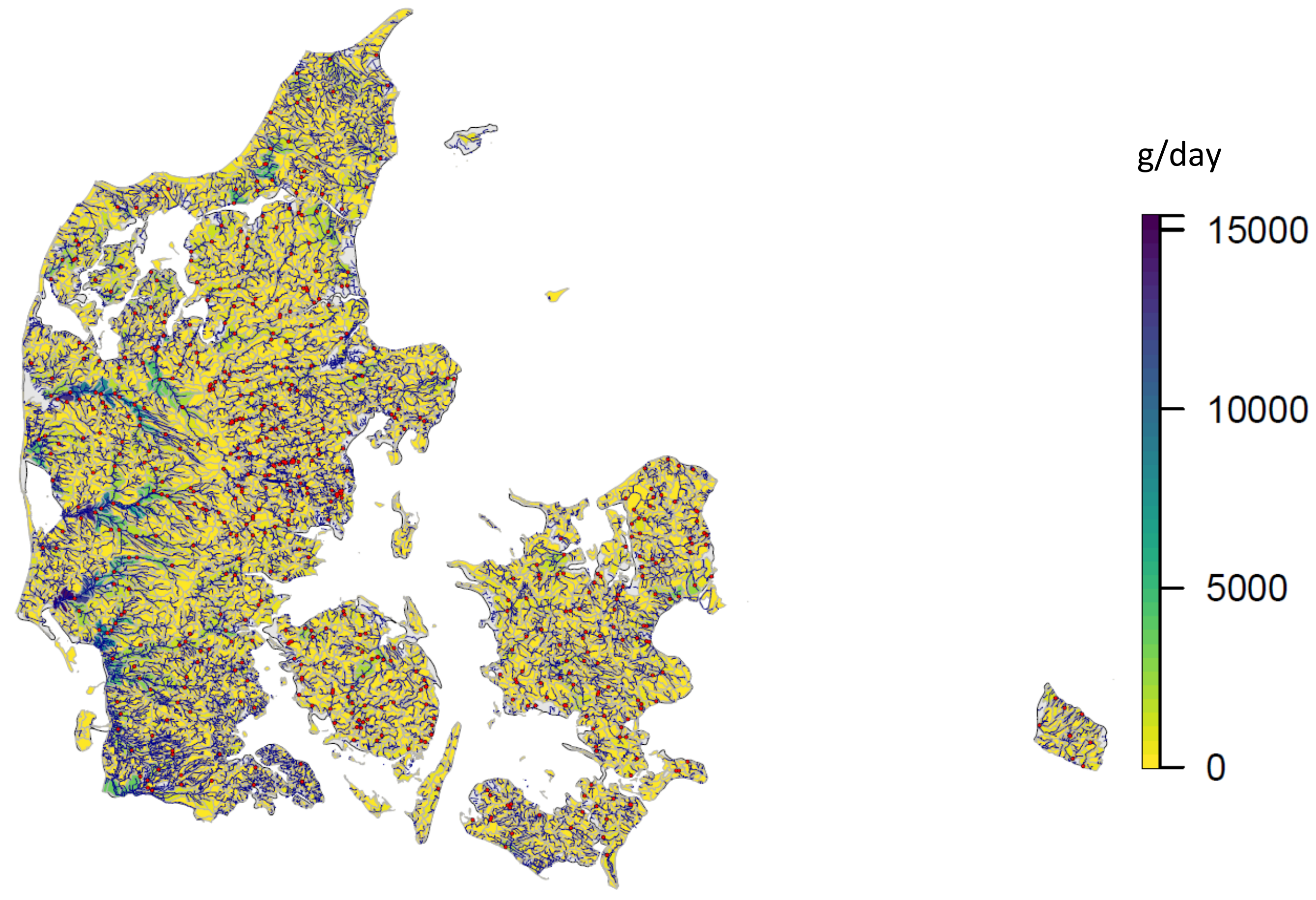

Figure 8. Estimated mass of nickel by multiplying the water volume flowing from the sub-catchments with the expected concentrations predicted by the single-source version.

## Two sources version

In the model, various explanatory variables can be utilized in the source model to predict sub-catchment outlet concentrations. From a list of potential explanatory factors at the sub-catchment level (e.g., soil pH, soil texture, groundwater concentration, degree of farming, human population), some of those can be selected based on knowledge regarding sources and transport mechanisms. In this paper the contribution from groundwater born source is included for illustration by assuming a two-source version where the mass input is divided into two main sources as groundwater born on one side and surface near water born on the other side. The groundwater source has primary mineralogical and thus natural origin, there is also anthropogenic contamination of groundwater, but that is considered limited compared to the mineralogical part for nickel, however. The surface near water born source is a mixture of many sources, however, in this simple version these sources are all lumped together as one unified mean concentration for all surface near water discharge into the sub-catchments yield the following source model:

$$X_{i,0} = Qs_i \tag{14}$$

$$X_{i,1} = Qg_i \cdot Cg_i \tag{15),}$$

where $Qs_i$ and $Qg_i$ are respectively the surface near and the groundwater discharge to sub-catchment $i$, as defined by Eq. 3. $Cg_i$ is the estimated groundwater reservoir concentration from where the groundwater is discharged to sub-catchment $i$. The $Qg_i$ value is estimated as the mean annual minimum monthly discharge calculated based on the period 2011-2023 and $Qs_i$ is estimated as the difference between the total generated water discharge to the sub-catchment and $Qg_i$. Such estimate is considered sufficiently valid for Danish conditions, where mid-summer water discharge is typically dominated by groundwater due to evaporation from the topsoil layers and plant uptake. The nickel concentration in the groundwater is estimated to be based on a combination of monitoring well data within the sub-catchments and groundwater magazine properties, such as pH and redox, for catchments with limited or no well data available (Supplementary). The resulting national mapping of the estimated groundwater concentration for nickel is displayed in Figure 9, disclosing some high concentration areas in the western part of Denmark.

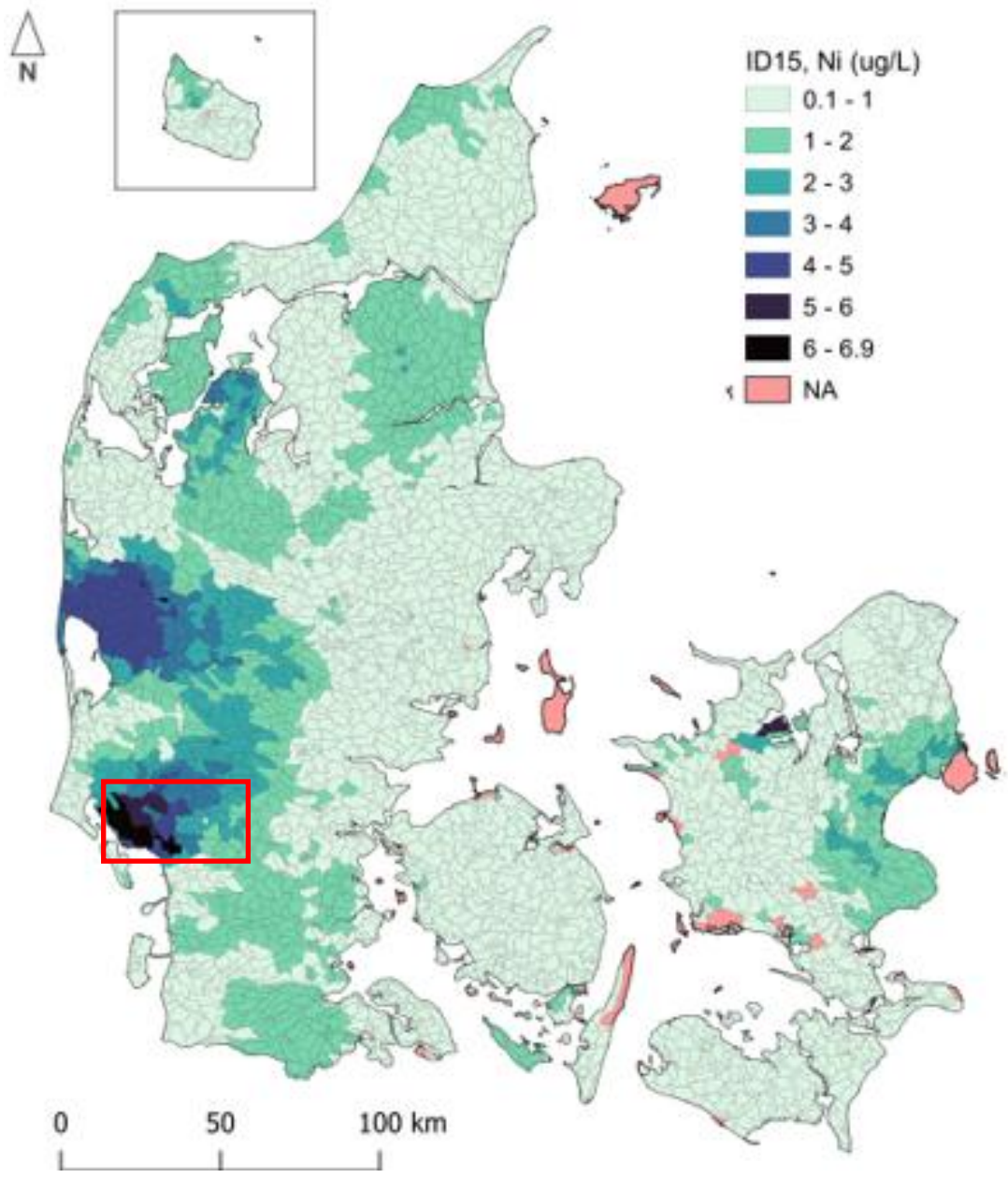

Figure 9. Estimated groundwater concentration of nickel based om well data and transfer function (Supplementary). The reed rectangle marks the area shown in Figure 10.

The intercept $\beta_0$ for the two-source version becomes the expected concentration of the generated discharge to the sub-catchment that is not groundwater (Eq. 14) and the equation for predictions becomes:

$$lC_i = \log\left(\frac{\sum_{j=1}^{N} C_j \cdot Q_j \cdot \Psi_{i,j}}{\beta_0} + X_{i,0} + \beta_1' \, X_{i,1}\right) - \log(Q_i) + \log(\beta_0) - \log\left(1 + \theta \, \frac{A_i}{Q_i}\right) + \varepsilon_i \tag{15),}$$

The general output is shown in Table 1 as Two-source.

Comparing the single and two-source version in Table 1 discloses similar estimates for the replicant standard deviation compared to the output in Table 1 ($\sigma_0$=0.440) and this also applies for the effect of years having a very small difference as $\sigma_Y$=0.124 (Single-source), $\sigma_Y$=0.121 (Two-source). This is expected because any improvement in the source-model by refining the sources will only improve the predictions for sub-catchments and thus leaving the variation between replicants unchanged. The same argument holds for effect of year as the catchment model (Eq. 9) for the two-source model does not include any yearly effects on the contributions of mass. Contrary, the estimated standard deviation for unexplained variation between sub-catchments has decreased from $\sigma_P$=0.563 (Single-source) down to $\sigma_P$=0.513 (Two-source). This decrease can be interpreted as follows: All the variation between sub-catchments as explained by the estimated groundwater source ($\sigma_{Pgr}^2$) is assumed to be part of the unexplained variation between sub-catchments in the single-source model ($\sigma_{P1}^2$) yielding the following relationship:

$$\sigma_{P1}^2 = \sigma_{Pgr}^2 + \sigma_{P2}^2 \qquad (16)$$

where $\sigma_{P1}$ is the unexplained standard deviation for the one source model ($\sigma_{P1}$=0.563), $\sigma_{Pgr}$ is the part of unexplained standard deviation in the single-source version that is caused by neglecting groundwater mass contribution, and $\sigma_{P2}$ is the unexplained standard deviation in the two-source version ($\sigma_{P2}$=0.513). So, it is possible to estimate $\sigma_{Pgr}$ by rearranging Eq. 16 and use the estimates for $\sigma_{P1}$ and $\sigma_{P2}$:

$$\sigma_{Pgr} = \sqrt{\sigma_{P1}^2 - \sigma_{P2}^2} \approx \sqrt{0.563^2 - 0.513^2} = 0.232 \qquad (17)$$

The Table 1 (Two-source version) shows that the retention is estimated smaller for the two-source version ($\theta$=25.28) compared to the single-source version ($\theta$=35.6), however, the estimate standard error in both tables for retention indicates high uncertainty to this estimate. The intercept has increased slightly from the single-source version ($\beta_0$=1.45 µg/l) to the two-source version ($\beta_0$=1.75 µg/l) this change in estimated concentration reflects the change in intercept reference. For the single-source version the intercept reference is the expected concentration in all volumes of water entering the sub-catchments, while for the two-source version the reference in the concentration is the expected concentration of the entering water volume not being groundwater. The $\beta_1'$ parameter in Table 2 is new compared to Table 1 and estimated to have a value of 0.529, and as seen in Eq. 8, this value needs to be multiplied by $\beta_0$ before it can interpreted as the coefficient for the mass contribution factor of nickel, which is illustrated in Table 2 as $\beta_1 = \beta_0\beta_1'$=0.93. It is interesting that this product is close to unity, as if the groundwater discharge to the sub-catchments is "correct" and the concentration estimates shown in Figure 8 are also "correct", then the Eq. 14 should be a "correct" estimate of the groundwater source to surface waters, and thus the $\beta_1$ values should be unity. This is highly interesting as a full-scale and data driven evaluation of the net results of the hyporheic zone, so in this case this zone has limited effect as sink for the interface between groundwater and surface water because $\beta_1 = 0.93 \approx 1$. Better understanding of this interface between the ground domain and surface water domain is stated by (Zhou et al., 2023) as an important topic for future research activities to support modeling metal exposure in surface waters.

The capability of the model to predict sub-catchment outlet concentrations seems to have improved by including the groundwater contribution as illustrated in Figure 9, which discloses the selected red rectangle cutting shown in Figure 9. This selection in Figure 8 is down to cover a higher estimated groundwater concentration level to the west and a lower estimated groundwater concentration level to the east. The selection is shown in Figure 10 for Single-source version and shows higher predicted concentration of the measured sub-catchments including catchments that have closely related water body, however, non-measured sub-catchments of typically smaller streams are not predicted high. This is in contrast to the left side of Figure 10, showing the two-source version, where high concentration is also predicted for the smaller streams and the predictions in general is following the groundwater concentration levels disclosed in Figure 10.

(a) Single-source version (b) Two-source version

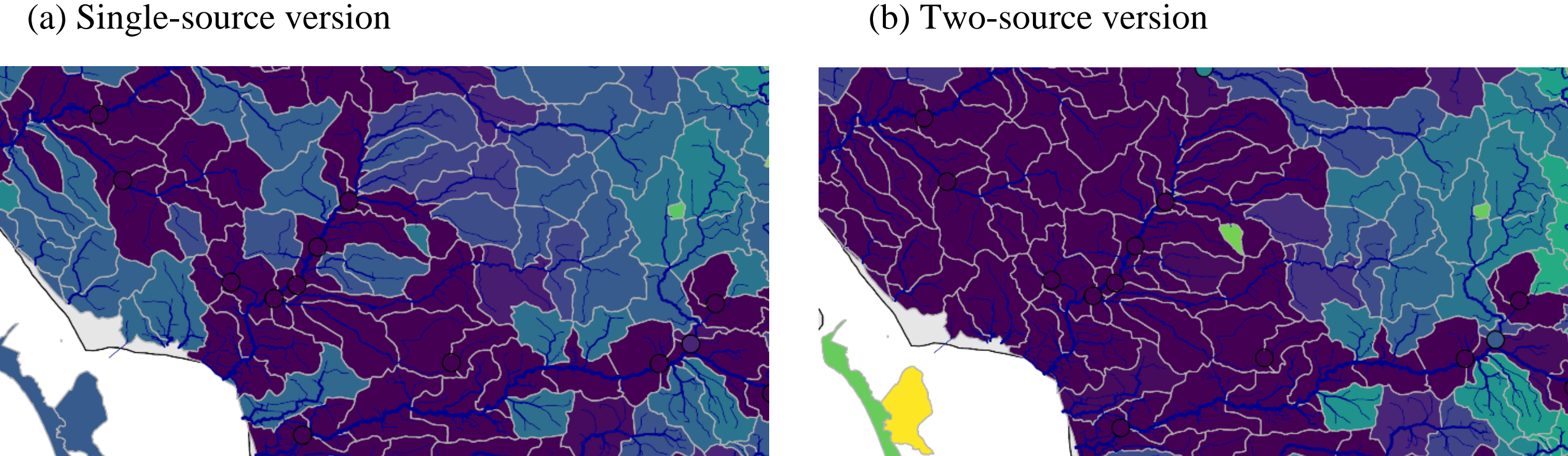

Figure 10. Predicted concentration for a part of Denmark (red rectangle shown in Figure 8): a): prediction made using the single-source version; b) Prediction made using the two-source version.

The two source version has divided the sources into two components and a key output must be to compare the estimated nationwide mass contribution of these two sources: Mass entering from groundwater is 91 kg/d, and mass from other sources is 252 kg/d. So, the sum of both sources is 341 kg/d, which deviates from the total estimate mass input of 387 using the single source model. However, the two-source version is obviously the most realistic model and thus the estimate of 341 kg/day most be seen at the best estimate under these conditions. A future refinement of the source model by adding more sources can improve the estimates of entering mass.

The single-source and two-source versions are further analyzed in Figure 11. For both versions data samples were sampled from 532 sub-catchments, representing 16% of the total 3350 sub-catchments. The model predicts the concentration for all 3350 sub-catchments; however, the measurements obtained from the sampled sub-catchments will have a particular impact on the prediction of concentration for the specific sub-catchment within which the samples have be collected. Additionally, the sampling not only affects the predictions for the sampled catchments but may also influence predictions for adjacent catchments, if they are hydrologically connected.

The primary objective in the following analysis is to validate the model's ability to predict concentrations in non-sampled catchments. To achieve this, a leave-*n*-out cross-validation procedure is applied to analyze the results from both the single-source and two-source versions. The procedure randomly selected 53 sub-catchments from the 532 sampled sub-catchments, for which the corresponding measurements were removed before recalculating the concentration predictions for the selected sub-catchments. The difference in predictions between the conditions with and without the inclusion of samples quantifies the model's performance in predicting concentrations for non-sampled catchments. The procedure is to divide the 532 measured sub-catchments randomly into 10 subsets. All the 10 sub sets are analyzed using a reduced data set where data from the selected 53 sub-catchment are excluded. This analysis will result in two predictions for 530 sampled sub-catchments: (1) a prediction using all samples; (2) A prediction, where the samples for the sub-catchment are excluded together with the samples for 52 randomly selected sub-catchments. The results of this analysis are shown in Figure 11, where each data point represents the concentration prediction (log-transformed) of a sampled sub-catchment. The x-axis displays the predictions including the measurements for that sampled sub-catchment, while the y-axis shows the predictions made neglecting these measurements.

As shown in Figure 11, the analysis is conducted for both the single-source and two-source version. When a sampled sub-catchment is located near other sampled sub-catchments, the predicted concentration may not be strongly influenced by the inclusion or exclusion of samples for the specific sub-catchment. In such cases, the data points in Figure 11 will lie close to the line $y=x$, indicated by the red dotted line. Although some data points cluster around this line for both the single-source and two-source models, most points are not closely aligned, particularly for the single-source version.

The horizontal line in Figure 11 represents the $\log(\beta_0)$ value from Table 1 (single-source), where $\beta_0 = 1.45$ µg/l is an estimate of the expected nickel concentration in water generated within a sub-catchment. If a non-sampled sub-catchment is not closely associated with any sampled sub-catchments, it will not be possible to predict a concentration higher than $\beta_0$, but only lower due to the retention as modeled by $\theta$. This explains why $\log(\beta_0)$ serves as an upper boundary for the majority of the data points in the single-source version of the model, as shown in Figure 11.

In case a sampled sub-catchment is located distant from other sampled sub-catchments then the predicted concentration will be strongly depended on whether the samples of this particular sub-catchment are included or not in the model, where exclusion of the samples will lead predictions close to a uniform nationwide prediction. So, especially the single-source version in Figure 11 divides the point cloud in two relations: (1) A relation close to $y=x$, which applies for sub-catchments that are close to other sampled sub-catchments; (2) a relation where the predicted concentration neglecting sampling of the specific sub-catchment is close, but under $\beta_0$, which applies sampled sub-catchments located distant from other sampled sub-catchments.

The two-source version shows distinct deviations from the single-source model. The division in the two relationships as seen for the single-source version is more unclear, however still indicated, and having more populated points around the $y=x$ line. A better capacity for the two-source version to perform predictions is also indicated by the $R^2$ value that improves from 0.28 to 0.37 between respectively the single-source and two-source versions.

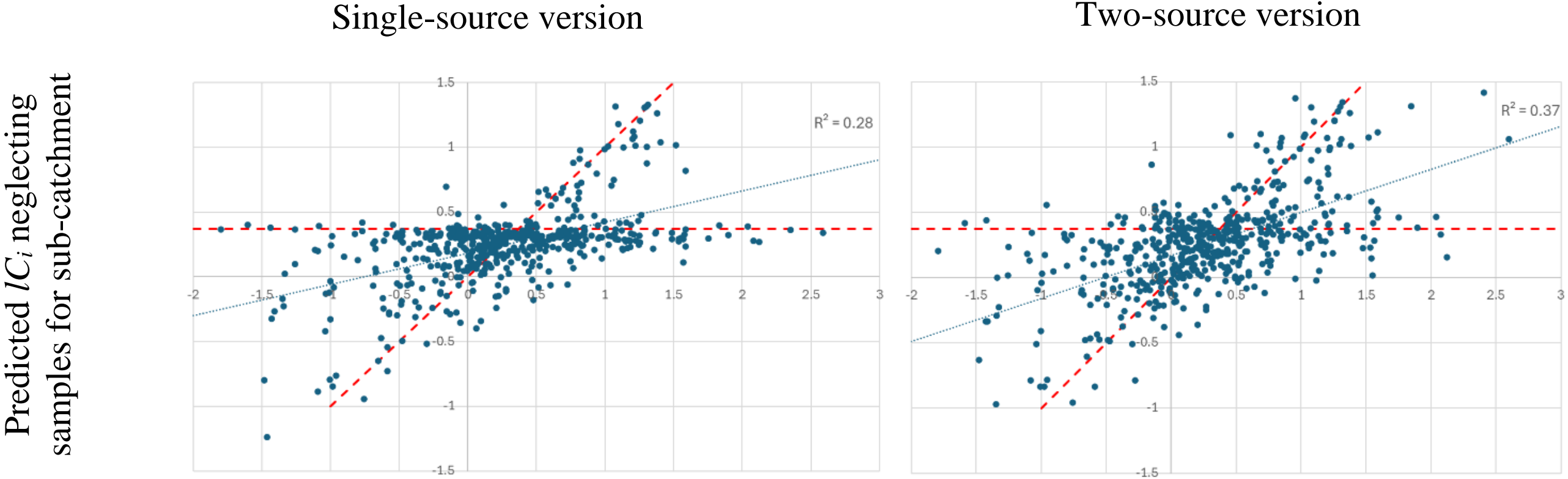

Predicted $lC_i$ including samples for sub-catchment

Figure 11. Comparison between the prediction of a sub-catchment depending on if the catchment has been sampled or not sampled. The one source version assumes the same concentration in water generated in sub-catchment, while the two-source version extend StatSurf with an estimated groundwater contribution.

# Discussion

The present study introduces a statistical modeling approach that integrates a monitoring dataset with a nationwide catchment model, incorporating the mass balance of a substance governed by water flow dynamics between sub-catchments. While the fundamental principle of the employed statistical model, based on a well-defined likelihood function, is not novel—having been in use for decades—the novelty of this work lies in the availability of a new solution package for solving large systems of statistical equations. This advancement extends the applicability of statistical models to complex catchment modeling scenarios, involving numerous sub-catchments within a water flow network, and considering the mass balance between sub-catchments in the inference of measurements.

Although this may seem like an academic achievement of limited application, it represents a paradigm shift in the capacity to develop and apply models based on monitoring data. Using this type of statistical modeling, it is possible to divide the correlation between measurements into two components: (1) Correlation that is trivial as a consequence of sampling downstream, more or less sharing the same water body; (2) Correlation that cannot be explained as resampling of the same water body and is thus subject to model development. Additionally, this approach allows for the estimation of replicate variation, thereby quantifying the fraction of measured concentration variation that is merely white noise within the sub-catchments and thus outside the reach of any model of the selected catchment setup. The catchment model can only address variability between sub-catchments, thereby reducing the variability between sub-catchments, which sets limits on its potential utility.

The case using Danish monitoring data of nickel illustrates how the addition of groundwater sources to the catchment model can yield reduced variability between sub-catchments, while replicant variability due to repeated measurements and annual differences remained unaffected. Traditionally, such variance component decomposition can only be performed with more well-defined and independent datasets compared to the complex monitoring data of surface water contaminants.

The model employs a left-censored likelihood function, and thus assuming all residuals to follow a single log-normal distribution. This means that all measurements below the detection limit are incorporated using the accumulated distribution, which is an appropriate method for handling measurements below detection limit when these residuals adhere to the same log-normal distribution as those above the detection limit. For metals like nickel, which typically exhibit a background level in most areas, the left-censored approach is both relevant and well-supported and this will also apply for nutrients. However, for substances with more sporadic exposure and no consistent background level like many of the organic pollutants, the left-censored approach may be insufficient. In such cases, the likelihood function could be extended to incorporate a zero-inflated model, a modification that could be straightforwardly implemented in future model versions.

The water flow and concentration measurements are assumed to be constant over time, with a yearly random effect on concentration. This assumption of constant flow and concentration within a year is particularly critical for regions with strong seasonal variations in hydrology. In the current version of the model, uncertainties due to such fluctuations primarily contribute to the standard deviation between replicates ($\sigma_O$). So, the upper limit of error due to such seasonal effects can be assessed based on the estimate of $\sigma_O$ and compared with the standard deviation between sub-catchments $\sigma_P$ and in case of nickel, where the two source version of StatSurf yields $\sigma_O = 0.440$ and $\sigma_P = 0.513$ is can be concluded that the temporal effect due to seasons may exist, but the variation between sub-catchments is larger, so there are still space for improvements of the source model by adding more sources to separate sub-catchments, also in this case where the seasonality is not taken into account.

As the model is a steady-state model, temporal dynamics affecting concentration tend to increase residual variability. This can be considered as a rather restrictive assumption especially for modelling nutrients, where the seasonality is important and in cases where the water discharge to the sub-catchments is exhibiting larger variations due to the seasonality. Nonetheless, there are numerous opportunities for future development of statistical models to address non-steady-state catchment models that can take seasonality into account in the mass balance. A non-steady state model will need to estimate a series of concentration levels for each sub-catchment outlet and thus extend the task for the equation solver. Nevertheless, in the case og nickel reported in this paper, the equation solver in RTMB only needs lesser than half a minute on a standard laptop, to solve all the equations for 3350 sub-catchments, so RTMB seems strong enough to solve the resulting equations also in non-steady state models.

The mass balance model presented in this paper assumes that the substance is in the water phase and passively transported by running water, which is a suitable assumption for many substances. This model is therefore not directly applicable to substances measured in biota, where the exchange of substances between sub-catchments does not solely follow the transport dynamics of running water either because the biota is not moving with the water like mussels or moves around independently on the water movements like fish. This limitation does not diminish the utility of the methodology introduced here, which combines monitoring data with model outputs through a well-defined likelihood function. Rather, it indicates the need for further development. For instance, an ecological model of biota development of biomass and movements in the catchment could be integrated as a “mass movement model”.

## Conclusion

The model approach described in this paper can utilize national wide monitoring data using catchment models by predicting concentration levels that fit measured concentrations including mass budges that can predict mass of contaminants entering marine areas and predict mass fractions retained or degraded in the surface water system. It is therefore possible to have maximal benefit from monitoring data to predict the influence spatial distribution of difference mass sources as defined by the catchment model including factors such as groundwater, agriculture, wastewater, etc.

The modelling apprach in this paper can take full benefit of the statistical properties of monitoring data to analyze the performance and parameter uncertainty of the monitoring model. This analysis is two-fold as it firstly set the model into contest of a variance component analysis, where the fraction of potential variability that can be explained by a "ideal" model is compared with the variability described by the model. Secondly the uncertainty is estimated for all parameters in both the statistical model and specifically in the monitoring model. These options are illustrated in the paper using nationwide Danish monitoring data of nickel and a simple monitoring model that divides the mass input into respectively a groundwater contribution and a surface near contribution.

This paper is a game-changer in relation to traditional catchment modelling, by combining comprehensive monitoring data sets with traditional catchment models. This is not a presentation of a new paradigm that will replace traditional modelling, but a method to facilitate improved synergy between traditional modelling and larger data sets. So, the model approach presented in this paper shall be seen as a first move that can help inspire other modelers to adjust and reformulate their own models to benefit better from monitoring data sets in their respective countries.

The model need a relatively short R code that runs efficiently on a standard computer and requires only basic computational skills to utilize R software packages and thereby employed in a systematic investigation of potential $X$ matrices by various research groups without the need for a server or other centralized system. Furthermore, the modeling approach can be used directly by staff at the Environmental Protection Agency (EPA), providing them with new opportunities to take advantage of the model's many capabilities.

## Acknowledgement

The Danish EPA (Miljøstyrelsen) is gratefully acknowledged for founding key activities that have led to this paper.

Supplementary materials

# Introduction

The purpose was to provide nationwide estimates of the potential nickel (Ni) input to surface water bodies from groundwater in Denmark. The typical Ni concentration (µg/L) contribution to each ID15 catchment was further used in the model *StatSurf* (see main text). The actual groundwater input of Ni to surface water is unknown, as the groundwater-surface water interactions have not been resolved at the national scale yet. Thus, here we describe the procedure we used to estimate them and the associated assumptions.

A simplified conceptual model is shown on Figure 1, where three groundwater bodies (abbreviated GVF here) with varying levels of Ni data: GVF1 has four well-screens with Ni data, GVF2 has only one well-screen with Ni data, and in GVF3 there are no well-screens, thus the Ni concentration is unknown.

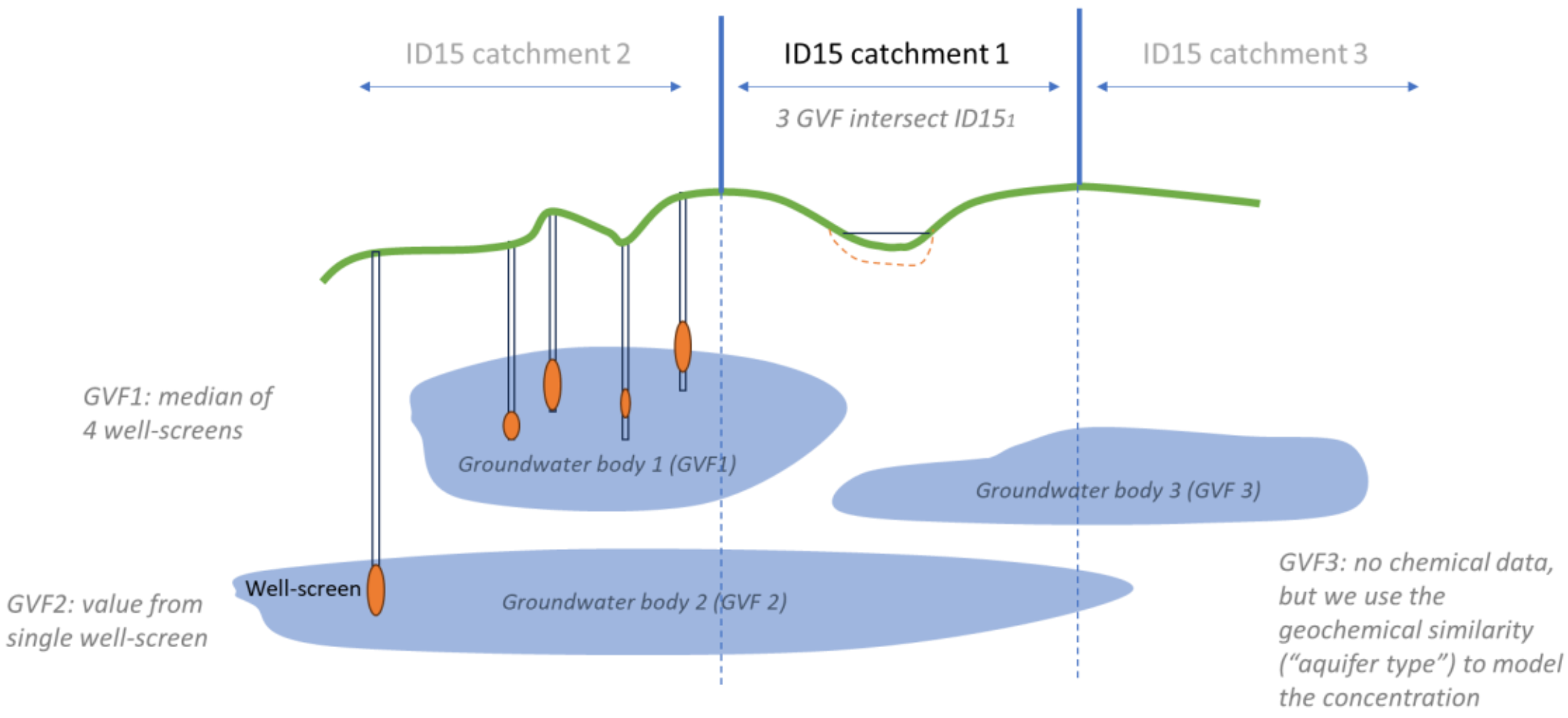

*Figure 1 Simplified conceptual model of the system showing the ID15 catchments at the surface and hypothetical groundwater bodies (GVF) intersecting spatially with the ID15 catchments. The GVF are known to be hydraulically connected to the ID15 catchments.*

Therefore, our methodology has three major steps (Figure 2), where the point observations of Ni at well-screens are transferred to the groundwater bodies and further to the ID15 catchments. We estimated in this order: 1) Ni concentration at the well-screen level (Mortensen et al. 2021), 2) Ni concentration at the GVFs connected to surface waters, 3) Ni input from GVF to the corresponding ID15 catchments (µg/L). More details are provided in the next section.

Our conceptual model does not reflect the geochemical processes potentially occurring in the riparian and hyporheic zones, i.e. the processes at the interface of groundwater and surface water, thus indirectly, we assumed that the groundwater concentrations of Ni would be unaltered along the groundwater flow-paths. Despite excluding many important geochemical processes, this assumption was deemed acceptable for the purposes of *StatSurf*, as the goal was to reflect the geographic variation in Ni concentrations in Danish groundwater, due to different geological, geomorphological, hydrological, and hydrogeochemical conditions in Denmark.

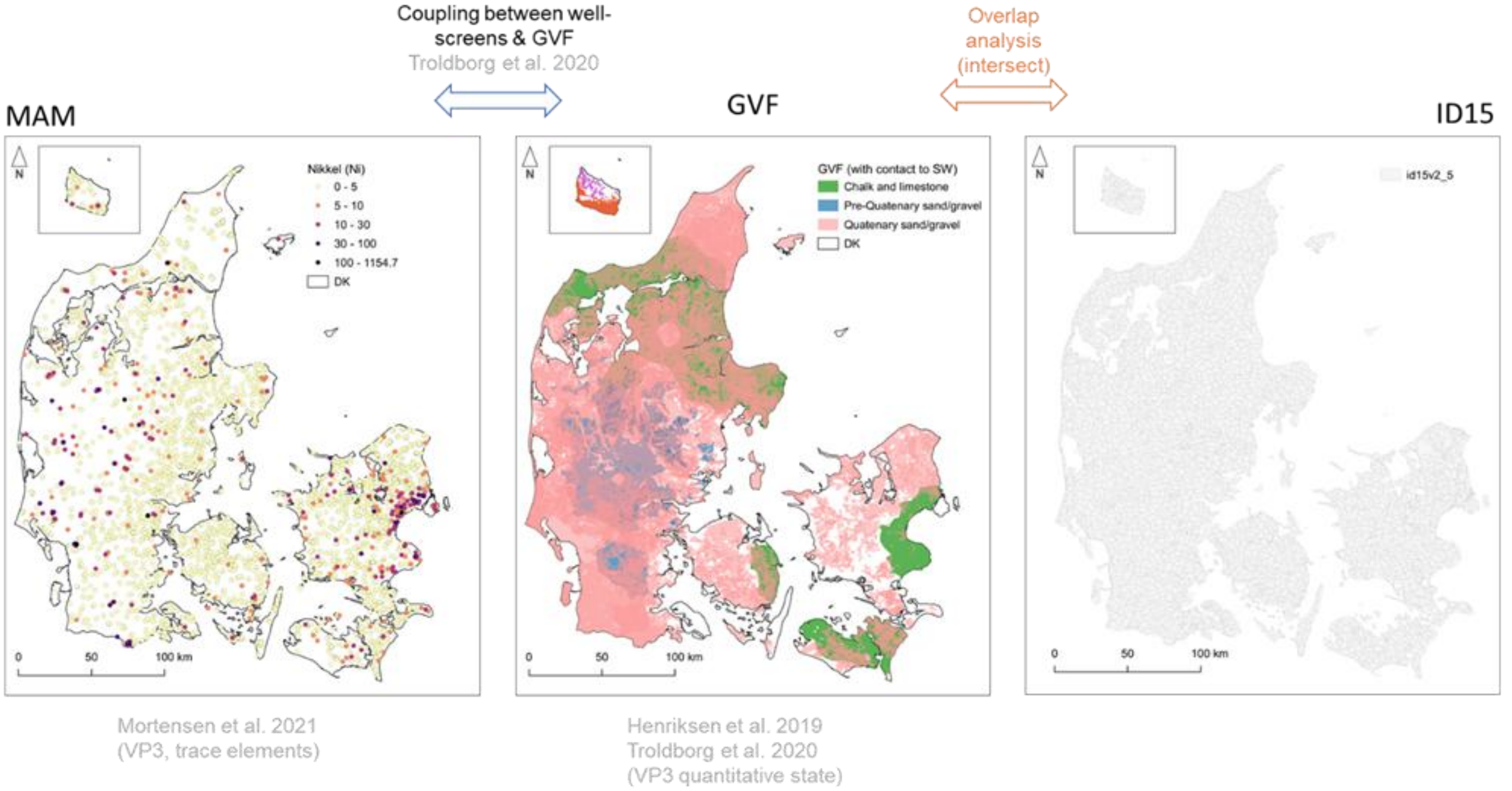

*Figure 2 Visualization of the three methodological steps: Ni concentrations at the well-screen level (left), groundwater body (mid), ID15 catchment (right); abbreviations: MAM – mean annual mean, GVF – groundwater body, ID15 – surface water catchments.*

# Method and data

## Workflow

The three-stepped methodology was formalized with the flow-chart in Figure 3. The first step (Figure 3) was a check whether the ID15 catchment intersected with any of the GVF in contact with surface waters (Table 1). For those ID15 that did not intersect with a GVF, the groundwater contribution of Ni was unknown, so "NA" (i.e. missing value) was saved for the specific ID15 catchment in the output file. If there was intersection, the second part of the algorithm was executed.

Second, for each GVF there was a check on how many well-screens with Ni data were associated with the specific GVF. For GVF with:

- > 2 well-screens with Ni data, a GVF-median was calculated based on the direct observations within the GVF, further referred as "*estimated typical value*".
- ≤ 2 well-screens with Ni data (i.e. not enough data), a median for the entire aquifer type was calculated instead. In this case we "*modeled"* the typical Ni concentration by transferring knowledge from other locations where there were direct observations. These other locations were selected based on similarity between the GVF without data and aquifers (a larger subsurface unit) that has similar conditions. The definition of aquifer types is provided in the next section.

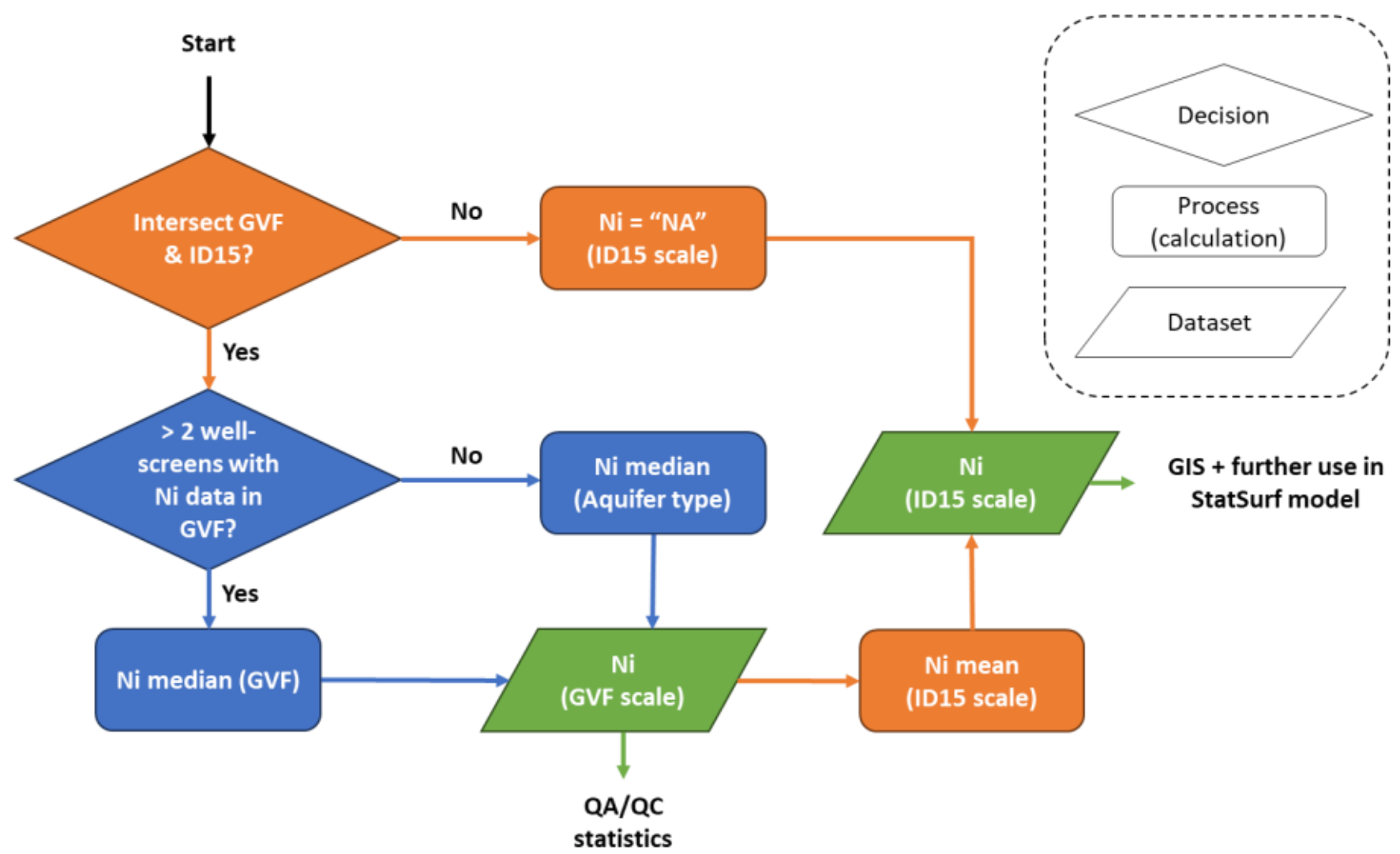

*Figure 3 Flow chart for transfering the nickel (Ni) concentration contribution from groundwater to the ID15 catchments; NA – missing (unknown) value; QA/QC – quality assurance and quality control procedures; GVF – groundwater body, GIS – geographic information system.*

*Table 1 Data and corresponding source.*

| Data | Reference |
| --- | --- |
| Mean annual mean concentration (MAM) of Ni in well-screens from the national monitoring network (GRUMO) and waterworks wells (BK) for the period 2000-2018 (incl. both years). Well-screen meta-data from the same source. | Mortensen et al. (2021) |
| Groundwater bodies (GVF) *in contact with surface waters* from the DK-model 2019 | Henriksen et al. (2019, 2021)<br>Troldborg et al. (2020) |
| Coupling between well-screens and GVF | Troldborg et al. (2020) |
| ID15 catchments, v.2 | Troldborg et al. (2015) |

Both the *estimated* and the *modelled* median Ni concentrations were then saved, so each GVF was assigned a single Ni value. This intermediate data-output was used for internal quality assurance and quality control (QA/QC) procedures before the third, last step of the algorithm was executed.

Last, we calculated a mean of the typical Ni at the GVF intersecting the ID15 and saved these values for the corresponding ID15 catchments in the data-output (Figure 3). In this step, three scenarios were possible, based on type of data used:

- mix of *modelled* and *estimated* typical Ni concentrations at the GVF level (Figure 4a)
- only *modelled* typical Ni concentrations at the GVF level (Figure 4b)
- only *estimated* typical Ni concentrations at the GVF level (Figure 4c).

These three scenarios reflect the varying degree of uncertainty in the Ni contribution to ID15 catchments; ID15 catchments where Ni concentration was based only on modelled data (no direct observations within the GVFs) are most uncertain, while those based only on direct observations are most certain.

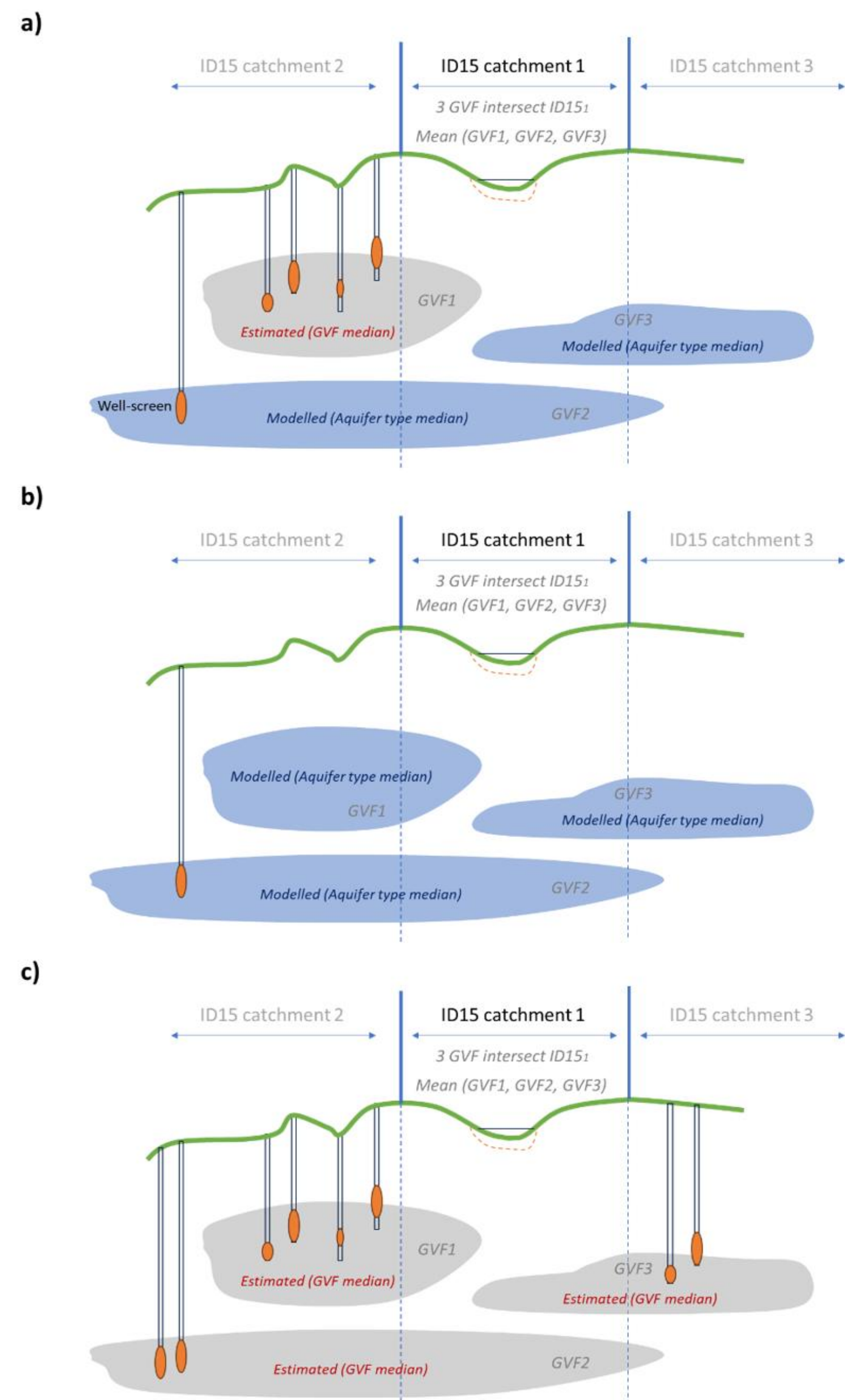

*Figure 4 Three scenarios for calculating the groundwater contribution to the ID15 catchments, based on: a) mix of modelled and estimated at the GVF level Ni concentrations; b) only modelled Ni concentrations; c) only estimated Ni concentrations.*

Figure 5 shows the number of GVF (from 0 to 13) intersecting the ID15 catchments. Here only GVF in contact with surface waters were considered.

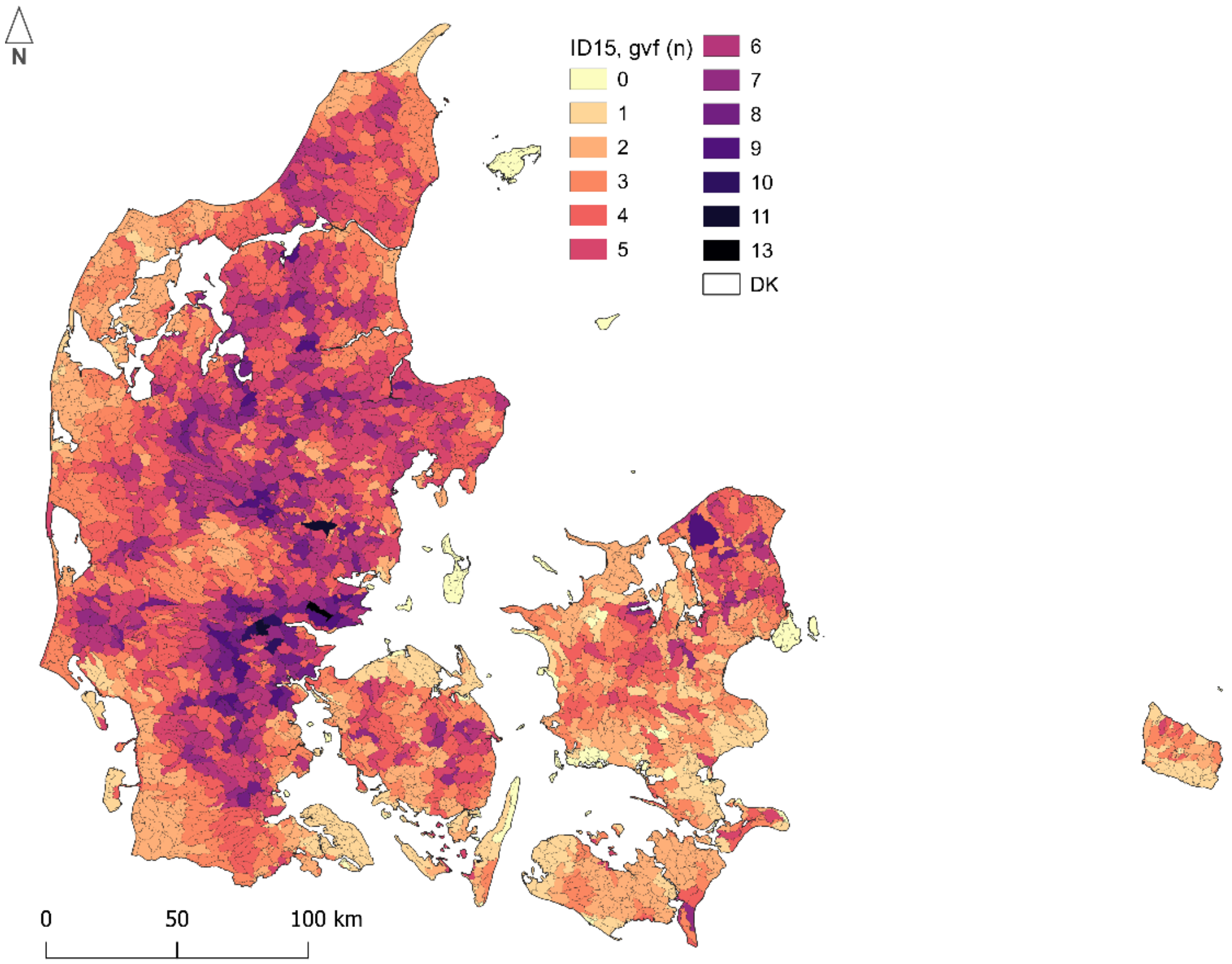

*Figure 5 Number of groundwater bodies in contact with surface waters (GVF) that intersect the ID15 catchments; the analysis was done in QGIS ("intersect")*

To summarize, the following calculations were made at different spatial levels, while executing this procedure (Figure 3):

- **ID15 catchment level**: mean of the typical Ni concentrations at the GVF intersecting a specific ID15. It was considered whether a weighted mean should be used instead, where the weight was the GVF volume, but we decided against it, because there was not enough evidence that larger GVF would contribute to the surface waters more than smaller GVF.
- **GVF level**: a typical Ni concentration was calculated, based on a median of the typical concentrations at the well-screens within the GVF or the aquifer type. The median was used because it is more robust against outliers and the non-normality of the Ni data. Two different calculations were done at this level, based on how many well-screens with data were there in the GVF:
    - If there were > 2 well-screens in the GVF – a median of Ni concentration at these well-screens was calculated, referred to as "estimated" Ni concentrations.
    - If there were ≤ 2 well-screens in the GVF – the typical Ni concentration for the specific aquifer type was used instead. It was modelled as the median of all well-screens that were linked to the specific aquifer type (see below for more details).
- **Well-screen level**: mean annual mean (MAM) Ni concentration for the period 2000-2018 (incl. both years) was from Mortensen et al. (2021). MAM was used because the Ni time-series were highly irregular (different sampling frequencies in different years) and this way each year had equal wight. This was done as part of the qualitative assessment for the third period of the River Basin Management Plan (RBMP 3).
- **In GIS**: an intersect between ID15 and GVF polygons to determine which GVF were overlapping spatially with which ID15 catchments. This was done only for GVF known to have contact with surface waters (Table 1).

## Defining aquifer type and groundwater body type

The definition of aquifer types follows the methodology developed for estimating natural background levels for RBMP 3 (Mortensen et al., 2021). The definition combines classification based on geology, geography, pH and redox conditions.

Both redox and pH have importance for the mobility/sorption of Ni. Both well-screens and GVF were classified accordingly.

The well-screens were classified, based on:

- **Redox conditions**: in reduced or oxic, based on nitrate concentration (≤ 2 mg/L and >2 mg/L).
- **pH:** low, neutral, high (≤ 6, 6-7, >7, respectively), based on the pH at the well-screen. This differs from Mortensen et al. (2021), as one additional class was added here for high pH (>7). The classes were redefined based on Figure 6.
- **Location and geology (nbl_unit)**, a combination of location (”dkms” – Sjælland, ”dkmf” – Fyn, ”dkmj” – Jylland, ”dkmb” – Bornholm) and geology (“ks” – Quaternary sand, “ps” – pre-Quaternary sand, “kalk” – carbonate aquifer, “uu” – various units on Bornholm).

Aquifer types were then defined based on the combination of redox, pH, and nbl_unit, such that Ni data from all well-screens within a specific aquifer type were grouped together, and median Ni concentration was calculated for each group (i.e. aquifer type).

Similarly, the GVF were also classified based on redox, pH, and nbl_unit, so they can be matched to aquifer type. The following was done to classify the GVF for:

- **Redox conditions:** the input data-source for the GVF (Table 1) contained information on the volumetric proportion of the GVF which is oxic (“OxProcent”), calculated with the National nitrogen model (Højberg, 2020). Thus, GVF with OxProcent >50% were classified as oxic, otherwise – **reduced**.
- **pH conditions**: The same pH classes were used as with the well-screen level. The GVFs have not been previously classified based on pH, so here we calculated a median pH at the GVF with QGIS zonal statistics, based on an IDW interpolated map from the pH at the well-screens (Figure 6). The interpolation was done at a coarse resolution (1km) reflecting the low-density of the available data (5,673 well-screens), also from Mortensen et al. (2021).
- **nbl_unit** was the same as above (well-screen level).

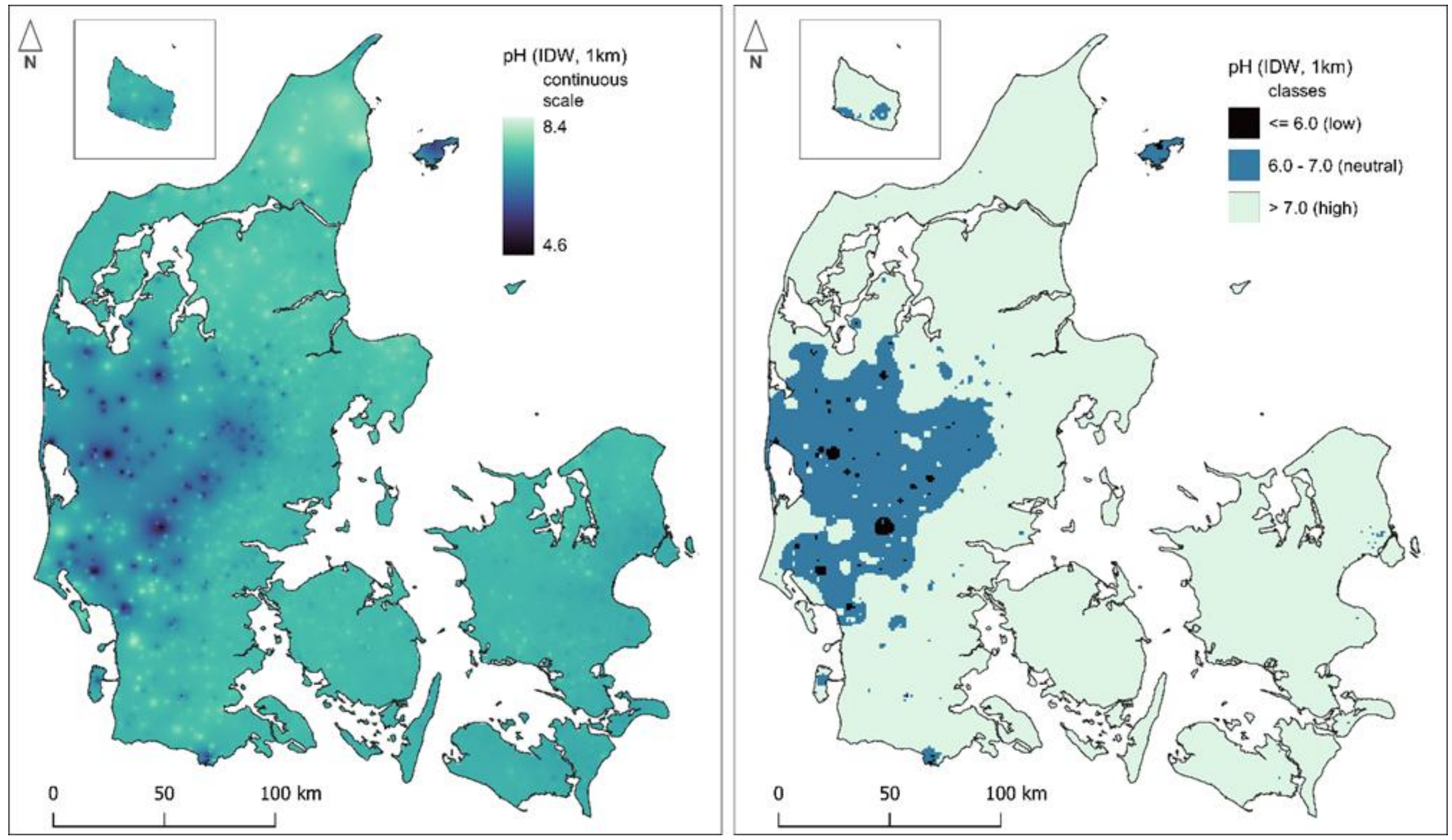

*Figure 6 Interpolated groundwater pH map (inverse distance weighted, IDW, 1km resolution) based on 10y-mean for 5,673 well-screens: left – continuous pH scale in the range 4.6–8.4, and right – same map, classified in the three pH classes (≤ 6, 6-7, >7). The groundwater with lowest pH is found in west and mid Jutland.*

# Results

## Classification of groundwater bodies (GVF) in contact with surface waters

In total more than a third (36%, n=740) of all groundwater bodies (GVF, n=2050) were in contact with surface waters. Most of these (85.7%) were surficial ("Terrænnær"); the rest were regional ("Regional") GVF. They were all relatively shallow, as the median (and IQR) depth was 4.6 m (0.67-13.9m) (based on mean GVF depth). Nearly 10% of the 740 GVF were layered. A 70% were in the first three (shallowest) Quaternary sand layers of (ks1, ks2, ks3), while 15 were in the carbonate aquifer layer of the national resource model.

Figure 7 and Figure 8 show the spatial extent of the GVF, classified by aquifer type (Table 2). In total there were 17 different aquifer types (Table 2).

*Table 2 Aquifer type classification of each groundwater body (GVF)*

| Aquifer type | Geography | Geology | pH (-) | NO3 (mg/L) | GVF (n) |
|---|---|---|---|---|---|
| dkmj_ks/Oxic/low pH | Jylland | Quaternary sand | ≤ 6 | > 2 | 1 |
| dkmj_ks/Reduced/low pH | Jylland | Quaternary sand | ≤ 6 | ≤ 2 | 1 |
| dkmb_uu/Oxic/high pH | Bornholm | Various | >7 | > 2 | 2 |
| dkmf_kalk/Reduced/high pH | Fyn | Carbonates | >7 | ≤ 2 | 2 |
| dkmf_ks/Oxic/high pH | Fyn | Quaternary sand | >7 | > 2 | 3 |
| dkms_kalk/Reduced/high pH | Sjælland | Carbonates | >7 | ≤ 2 | 4 |
| dkmj_kalk/Reduced/high pH | Jylland | Carbonates | >7 | ≤ 2 | 9 |
| dkmj_ks/Oxic/neutral pH | Jylland | Quaternary sand | 6-7 | > 2 | 11 |
| dkmb_uu/Reduced/high pH | Bornholm | Various | >7 | ≤ 2 | 17 |
| dkmj_ks/Reduced/neutral pH | Jylland | Quaternary sand | 6-7 | ≤ 2 | 25 |
| dkmj_ps/Reduced/neutral pH | Jylland | Pre-Quaternary sand | 6-7 | ≤ 2 | 25 |
| dkmj_ps/Reduced/high pH | Jylland | Pre-Quaternary sand | >7 | ≤ 2 | 32 |
| dkms_ks/Oxic/high pH | Sjælland | Quaternary sand | >7 | > 2 | 47 |
| dkmf_ks/Reduced/high pH | Fyn | Quaternary sand | >7 | ≤ 2 | 75 |
| dkmj_ks/Oxic/high pH | Jylland | Quaternary sand | >7 | > 2 | 97 |
| dkms_ks/Reduced/high pH | Sjælland | Quaternary sand | >7 | ≤ 2 | 173 |
| dkmj_ks/Reduced/high pH | Jylland | Quaternary sand | >7 | ≤ 2 | 216 |

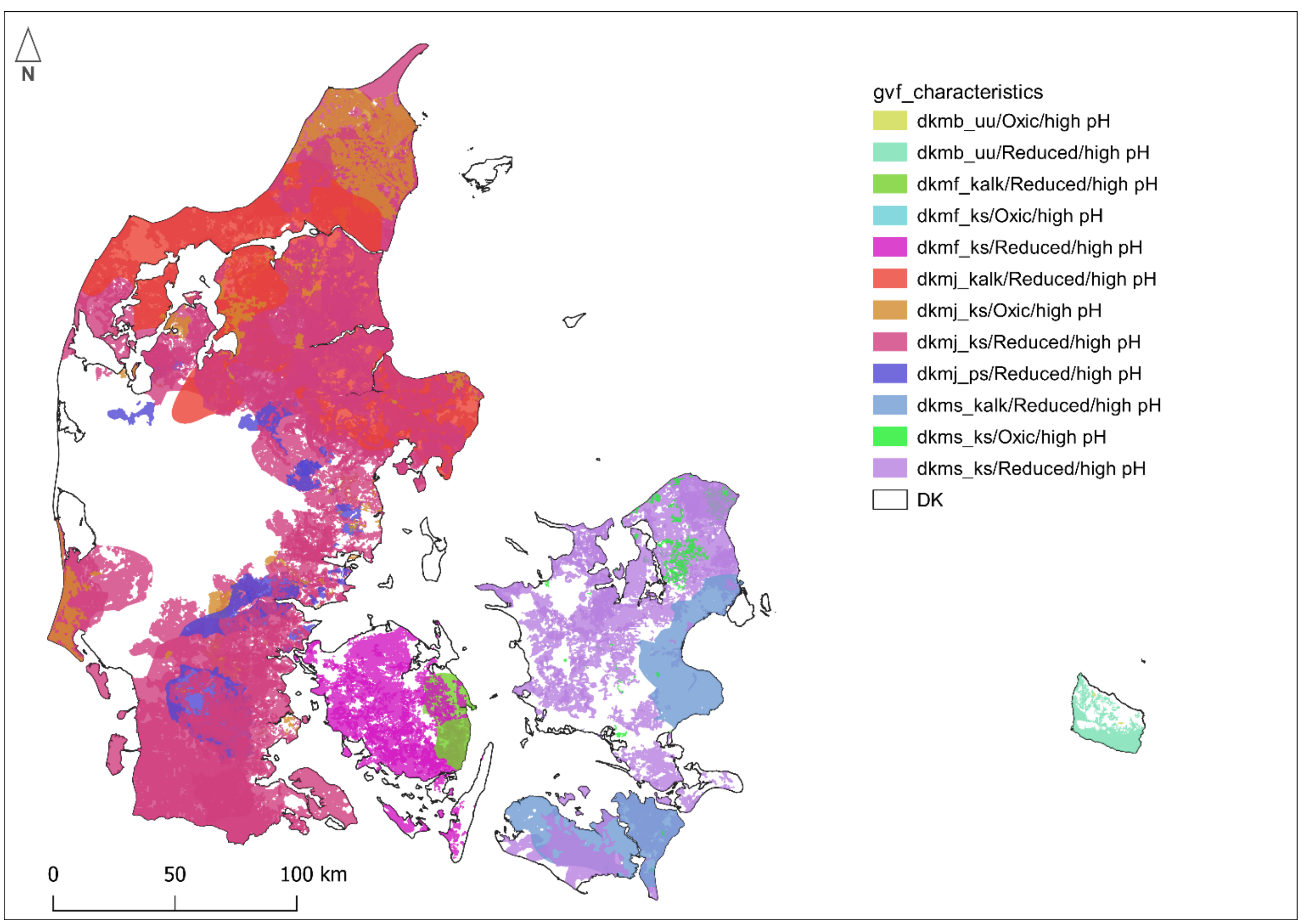

*Figure 7 Classification of groundwater bodies (GVF) in contact with surface waters. Here only the GVF with* ***high pH*** *(> 7) are shown, see Figure 8 for the rest; transparency is used so the spatial overlap of the GVF can be visualized. GVF can overlap both laterally and in depth (located beneath each other).*

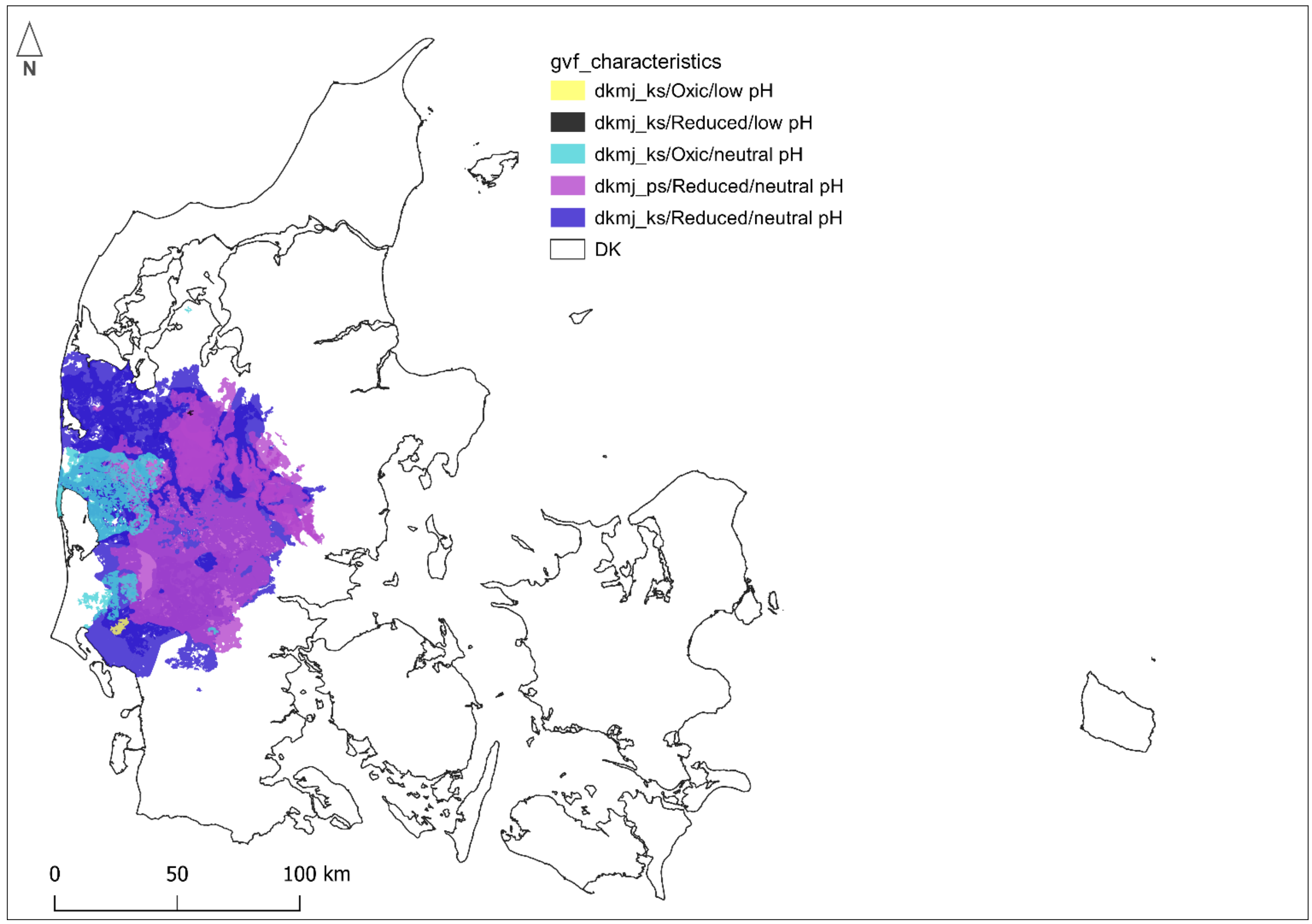

*Figure 8 Classification of groundwater bodies (GVF) in contact with surface waters. Only the GVF with* ***low or neutral pH*** *(≤ 6 and 6-7, respectively) are shown here, see Figure 7 for the rest; transparency is used so the spatial overlap of the GVF can be visualized. GVF can overlap both laterally and in depth (located beneath each other).*

## Ni concentrations at the well-screen level

In total, there were 5672 well-screens with Ni data. These well-screens were associated with 252 GVF in contact with surface water (GVF), and the Ni concentration at 179 GVF could be directly estimated from observations within the GVF. The rest were modelled or assigned NA, according to the described methodology.

The nationwide distribution of Ni concentration, stratified by pH and redox (Figure 9) shows that there was a well-defined difference in the Ni distributions for different combinations of pH and redox. The highest Ni concentrations were found in oxic groundwaters with low pH, while the lowest Ni were in reduced groundwaters with high pH.

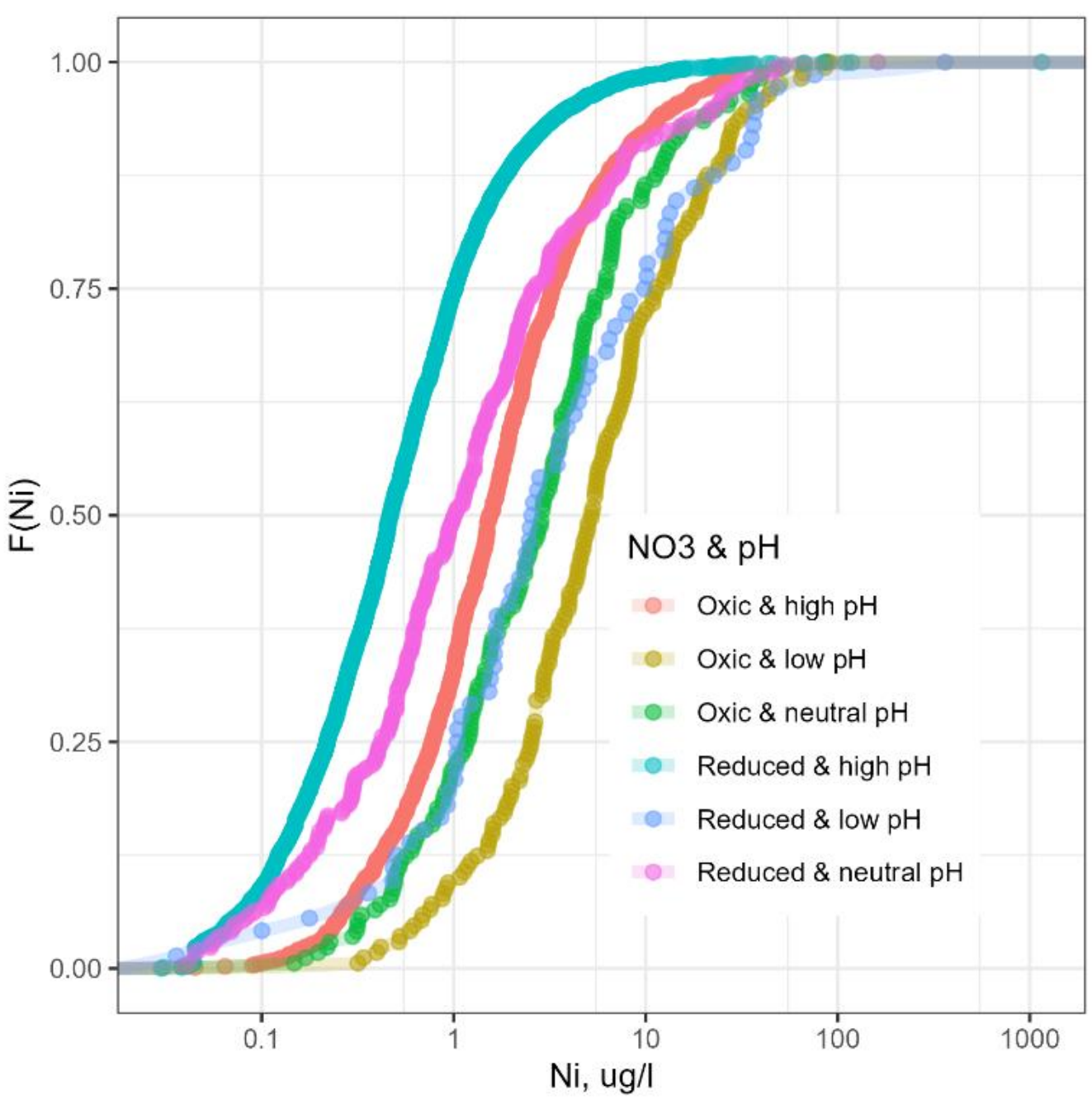

*Figure 9 Empirical cumulative distribution function for Ni in Denmark; each point is a mean annual mean (the MAM) concentration at a well screen, i.e. each well-screen is represented once here (data from Mortensen et al., 2021).*

## Ni concentrations at the groundwater body (GVF) level

The GVF with modelled typical Ni concentration are visualized on Figure 10. They were much shallower, with a median depth of 3.0 m (IQR = 0.2–7.6m), while the median depth for the GVF with direct Ni observations was 18.4m (IQR 8.1–34.8m). This shows that the GVF near the terrain, which are most probably affected by anthropogenic activities, were largely without direct Ni observations. Using *modelled* concentrations based on the aquifer types would then most probably result in underestimation of the typical Ni concentration in those GVF. The lack of direct observations reflects the pre-selection of well-screens for the purposes of setting natural background levels (Mortensen et al., 2021), where all shallow wells monitoring point-source pollution were excluded from the dataset.

There was no relation between the typical Ni concentration (i.e. median Ni) and the number of well-screens with Ni data within the GVF or aquifer type (Figure 10).

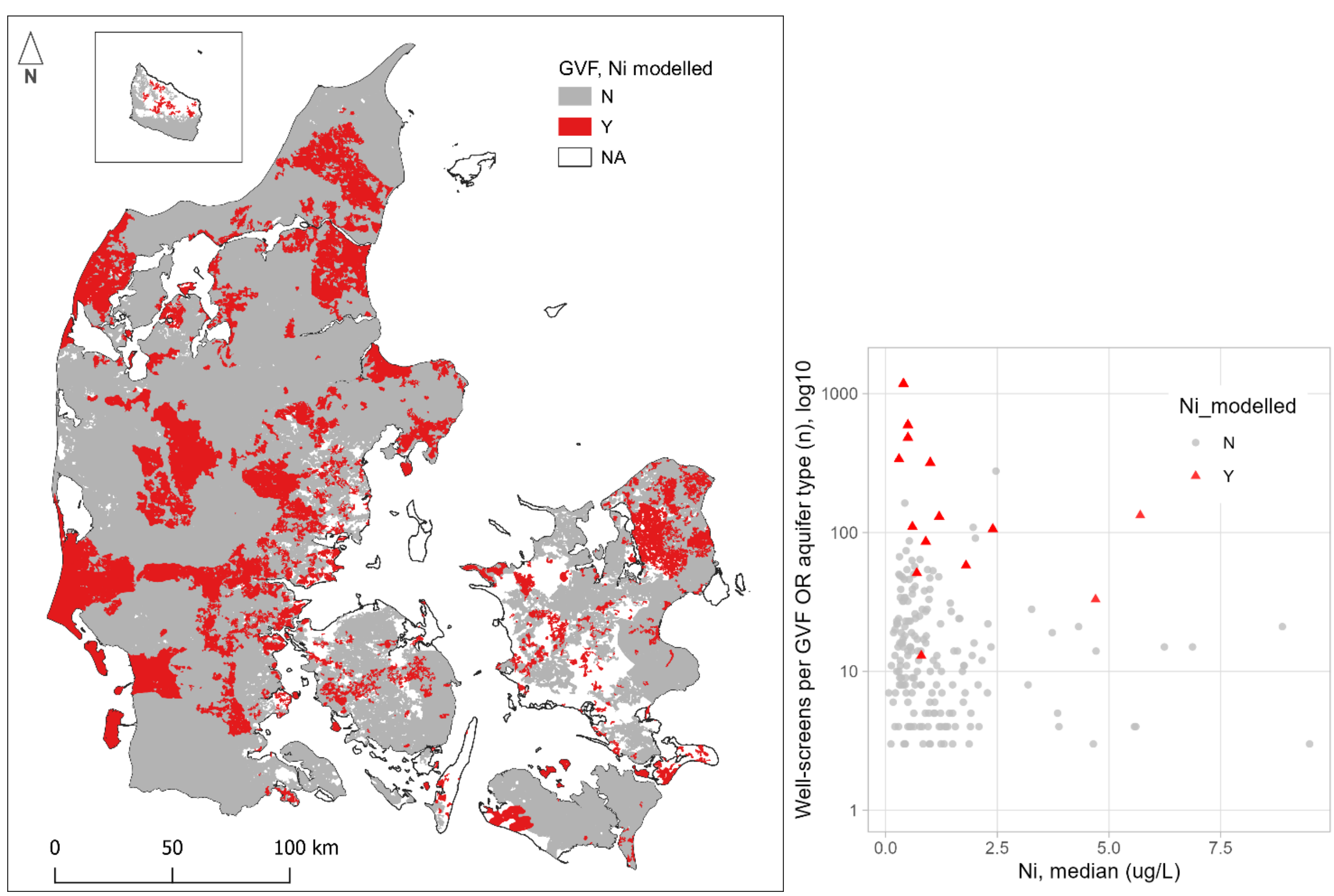

*Figure 10 left: Groundwater bodies (GVF), where the typical Ni concentration were modelled (Y – yes, N – no), based on aquifer type (if "N", the Ni concentrations were estimated based on direct observations from at least two well-screens within the GVF); right: median Ni concentration at the GVF (or aquifer), as a function of number of well-screens with Ni data per GVF (or aquifer type for the red triangles);*

## Groundwater contribution of Ni to the ID15 catchments

The groundwater contribution with Ni to the ID15 catchments in Denmark is visualized on Figure 11. The majority of the ID15 catchments have groundwater Ni contribution < 1 µg/L with higher Ni concentrations in western Jutland (up to 6.9 µg/L), largely due to low groundwater pH. The Ni contribution in Eastern Zealand is slightly higher that other parts of the country (1-3 µg/L), which is considered to be caused by drinking water abstraction, imposing a more dynamic groundwater potential(Larsen and Postma, 1997)(first rapid lowering of groundwater table/potential, then recovery after lowering the pumping rates). The data from Figure 11 was further used in the *StatSurf* model*.*

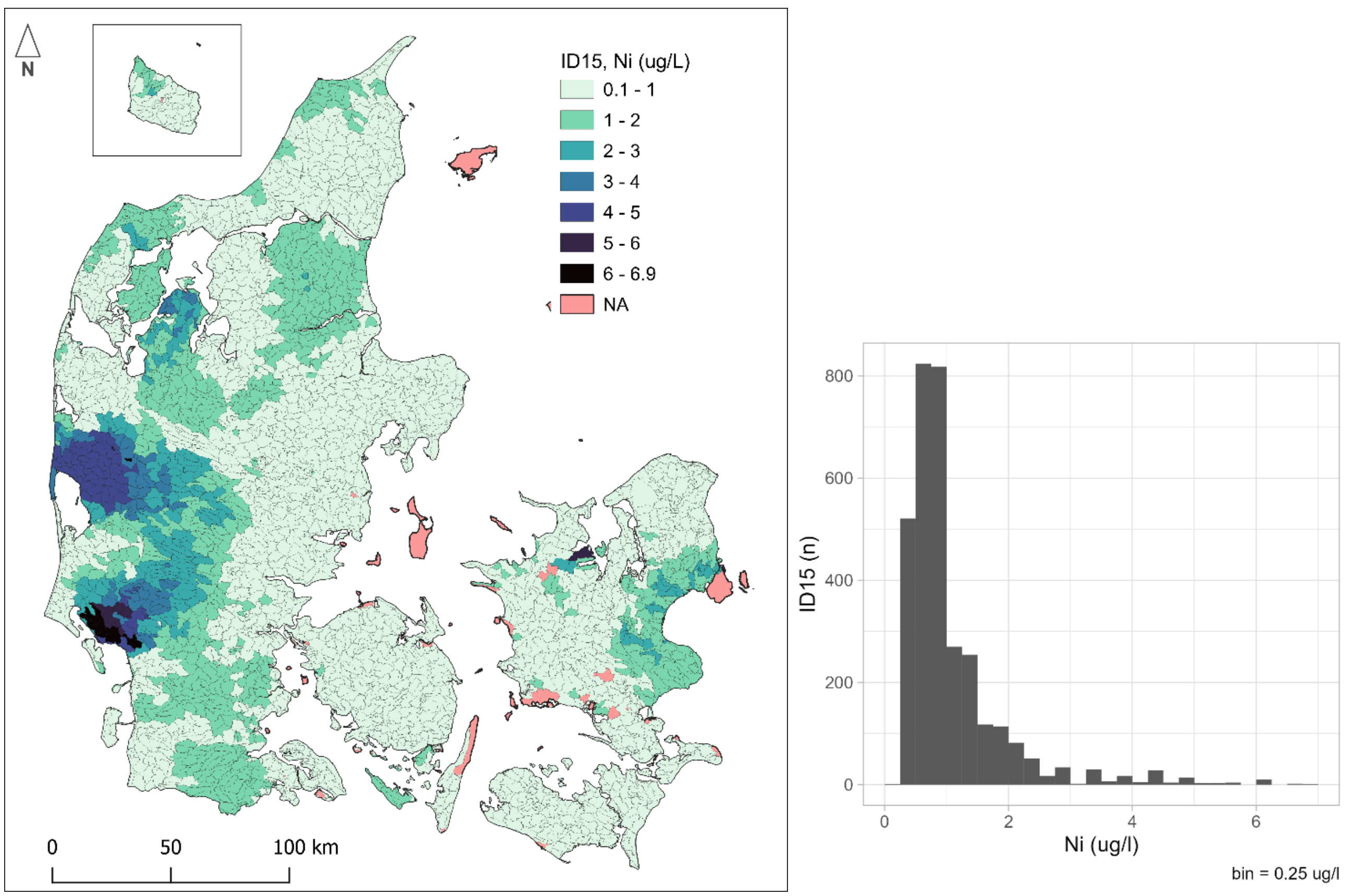

*Figure 11 left: map with the groundwater contribution with Ni (µg/L) to each of the surface water catchments (ID15 catchments). Right: histogram of ID15 catchments and their typical groundwater Ni concentrations (µg/L)*

Table 3 shows the number of ID15 catchments by method of calculating the groundwater Ni contribution, and the associated uncertainty level. The 117 ID15 catchments with unknown Ni contribution, are those which had no intersection with GVF or were outside of the modelling domain of the national water resources model.

*Table 3 Number of ID15 catchments with Ni contribution, by method based on the level of direct observations within the GVF. In total there were 3351 ID15 catchments; See Figure 4 for definition of "estimated", "modelled", "mix"*

| **Method (data-availability)** | **Estimated (n,%)** | **Modelled (n)** | **Mix (n)** | **NA (n)** |
|---|---|---|---|---|
| **Ni concentration** | 1419 (42.3%) | 141 (4.2%) | 1674 (50.0%) | 117 (3.5%) |
| **Uncertainty** | High | Low | Moderate | Unknown |